\documentclass[format=acmsmall, natbib]{acmart}

\usepackage{booktabs} 
\usepackage{bbold}

\makeatletter
\def\NAT@spacechar{~}
\makeatother

\usepackage[ruled]{algorithm2e} 

\SetAlFnt{\small}
\SetAlCapFnt{\small}
\SetAlCapNameFnt{\small}
\SetAlCapHSkip{0pt}
\IncMargin{-\parindent}

\usepackage{color}
\usepackage{beramono}
\usepackage{xspace}

\usepackage{paralist}  

\usepackage{booktabs}
\usepackage{tabularx}
\usepackage{multirow}
\usepackage{multicol}

\newcommand{\vect}[1]{\mathbf{#1}}  

\newcommand{\R}{\mathbb{R}}

\newcommand{\ngram}[0]{$n$-gram\xspace}
\newcommand{\ngrams}{$n$-grams\xspace}

\newcommand{\id}[1]{\lstinline{#1}} 
\newcommand{\codevec}[1]{\mathbb{#1}}  
\usepackage{listings}
\newcommand{\etal}{\hbox{\emph{et al.}}\xspace}
\newcommand{\eg}{\hbox{\emph{e.g.}}\xspace}
\newcommand{\ie}{\hbox{\emph{i.e.}}\xspace}

\newcommand{\etc}{\hbox{\emph{etc.}}\xspace}
\newcommand{\viz}{\hbox{\emph{viz.}}\xspace} 
\newcommand{\vs}{\hbox{\emph{vs.}}\xspace}

\begin{document}

\newcommand{\ourtitle}{A Survey of Machine Learning for Big Code and Naturalness}

\markboth{Allamanis \etal}{\ourtitle}

\title{\ourtitle}
\author{Miltiadis Allamanis}
\affiliation{%
 \institution{Microsoft Research}
 \city{Cambridge}
 \country{United Kingdom}}
 
\author{Earl T. Barr}
\affiliation{%
  \institution{University College London}
  \city{London}
  \country{United Kingdom}}
\author{Premkumar Devanbu}
\affiliation{%
  \institution{University of California, Davis}
  \state{California}
  \country{USA}}
\author{Charles Sutton}
\affiliation{%
  \institution{University of Edinburgh}
  \city{Edinburgh}
  \country{United Kingdom}}
\affiliation{%
    \institution{The Alan Turing Institute}
    \city{London}
    \country{United Kingdom}}

\begin{abstract}
Research at the intersection of machine learning, programming languages, and
software engineering has recently taken important steps in proposing learnable
probabilistic models of source code that exploit code's abundance of patterns.
In this article, we survey this work. We contrast programming
languages against natural languages and discuss how these similarities and
differences drive the design of probabilistic models.  We present a taxonomy
based on the underlying design principles of each model and use it to navigate
the literature.  Then, we review how researchers have adapted these models to
application areas and discuss cross-cutting and application-specific challenges
and opportunities.
\end{abstract}

%
%


%
%




\thanks{
This work was supported by Microsoft Research Cambridge through
its PhD Scholarship Programme. M. Allamanis, E. T. Barr, and C. Sutton are supported by
the Engineering and Physical Sciences
Research Council [grant numbers EP/K024043/1, EP/P005659/1, and EP/P005314/1].
P. Devanbu is supported by the National Research Foundation award number 1414172.

Author's addresses: M. Allamanis, Microsoft Research, Cambridge, UK;
E.T. Barr, University College London, UK;
P. Devanbu, UC Davis, CA, USA;
C. Sutton, University of Edinburgh, UK and Alan Turing Institute, London, UK;
}

\maketitle

\renewcommand{\shortauthors}{Allamanis \etal}

\section{Introduction}
\label{sec:intro}
Software is ubiquitous in modern society. Almost every aspect of life, including
healthcare, energy, transportation, public safety, and even entertainment, depends
on the reliable operation of high-quality software. 
Unfortunately, developing software is
a costly process: software engineers need to tackle the inherent complexity of
software while avoiding bugs, and still  delivering highly functional
software products on time. There is therefore an ongoing demand for innovations in
software tools that help  make software more reliable and maintainable.
New methods are constantly sought, to reduce the complexity of software and help
engineers construct better software. 

Research  in this area has been dominated by the \emph{formal}, or \emph{logico-deductive}, approach.  
Practitioners of this approach hold that, since software is constructed in mathematically well-defined
programming languages, software tools can  be conceived in purely formal terms. The design of software
tools is to be approached using formal methods of definition, abstraction, and deduction. Properties of
tools thus built should be proven using rigorous proof techniques
such as induction over discrete structures. This  logico-deductive approach has tremendous appeal in programming languages research, as it holds the promise of proving facts and properties of the program.  
Many elegant and powerful abstractions, definitions, algorithms, and proof techniques have been developed,
which have led to important practical tools for program verification, bug finding, and refactoring~\cite{bessey2010few, cousot2005astree, clarke2003behavioral}. 
It should be emphasized that these are theory-first approaches. Software constructions are viewed 
primarily as mathematical objects, and when evaluating software tools built using this approach, the
elegance and rigor of definitions, abstractions, and formal proofs-of-properties are of dominant concern.
The actual varieties of \emph{use} of software constructs, in practice, become relevant later, in 
case studies, that typically accompany presentations in this line of work. 

Of late, another valuable resource has arisen: the large and growing body of successful, widely used,
open-source software systems. Open-source software systems such as Linux, MySQL, Django, Ant, and
OpenEJB have become ubiquitous. 
These systems publicly expose not just source code, but also meta-data
concerning authorship, bug-fixes, and review processes. 
The scale of available data is massive: billions of tokens of code and millions of
instances of meta-data, such as changes, bug-fixes, and code reviews (``big code''). 
The availability of ``big code'' suggests a new, data-driven approach 
to developing software tools: why not let the statistical distributional
properties, estimated over large and representative software corpora,  also influence the design
of development tools?  Thus rather than
performing well in the worst case, or in case studies,  our tools can perform well in \emph{most cases},
thus delivering greater advantages in expectation. 
The appeal of this approach echoes that of  earlier work in computer architecture: Amdahl's law~\citep{amdahl1967validity}, for example,
tells us to focus on the common case.  This motivates a similar hope for development tools,
that tools for software development and program analysis can be improved
by focusing on the common cases 
using a fine-grained estimate of the statistical distribution of code.
Essentially, the hope is that  analyzing the text of thousands of well-written software
projects can uncover patterns that partially characterize software that is reliable,
easy to read, and easy to maintain.

The promise and power of machine learning rests on its ability to generalize
from examples
and handle noise.  To date, software engineering (SE) and programming languages (PL)
research has largely focused on using machine learning (ML)
techniques as black boxes to replace heuristics and find features, sometimes
without appreciating the subtleties of the assumptions these techniques make.  A
key contribution of this survey is to elucidate these assumptions and their
consequences.  Just as natural language processing (NLP) research changed focus from 
brittle rule-based expert systems that could not handle the diversity of real-life data 
to statistical methods \cite{jurafsky2000speech}, SE/PL
should make the same transition, augmenting traditional methods that consider
only the formal structure of programs with information about the statistical
properties of code.

\paragraph{Structure}
First, in \autoref{sec:hypothesis}, we discuss the basis of this area, which we call
the ``naturalness hypothesis''. We then review recent work on machine learning methods
for analyzing source code, focusing on probabilistic models, such as $n$-gram language models
and deep learning methods.\footnote{It may be worth 
pointing out that deep learning and probabilistic
modeling are \emph{not} mutually exclusive. Indeed,
many of the currently most effective methods for
language modeling, for example, are  based on deep learning.} 
We also touch on other types of machine learning-based
source code models, aiming to give a broad overview of the area, to explain the
core methods and techniques, and to discuss applications in
programming languages and software engineering.
We focus on work that goes beyond a ``bag of words'' representation of code, 
modeling code using sequences, trees, and continuous representations.
We describe a wide range of emerging applications, ranging from recommender systems,
debugging, program analysis, and program synthesis. The large body of work on 
semantic parsing~\citep{neubig2016survey}, is not the focus of this survey but we
include some methods that output code in general-purpose programming languages (rather
than carefully crafted domain-specific languages).
This review is structured as follows. We first discuss the different characteristics
of natural language and source code to motivate the
 design
decisions involved in machine learning models of code (\autoref{sec:codevsnl}). We then
introduce a taxonomy of
probabilistic models of source code (\autoref{sec:probmodels}). Then we describe
the software engineering and programming language applications of probabilistic
source code models (\autoref{sec:applications}). Finally, we mention a few overlapping
research areas (\autoref{sec:related}), and 
we discuss challenges and interesting future directions (\autoref{subsec:futuredirs}).

\paragraph*{Related Reviews and other Resources} There have been short reviews
summarizing the progress and the vision of the research
area, from both software engineering \citep{devanbu2015new}
and programming languages perspectives
\citep{bielik2015programming,yahav2015programming}.
However, none of these articles can be considered
extensive literature reviews, which is the purpose of this work.
\citet{ernst2017natural} discusses promising areas of applying natural language
processing to software development, including error messages,
variable names, code comments, and user questions.
Some resources, datasets and code can be found
at \url{http://learnbigcode.github.io/}. An online version of
the work reviewed here --- which we will keep up-to-date
by accepting external contributions --- can be found at \url{https://ml4code.github.io}.


\section{The Naturalness Hypothesis}
\label{sec:hypothesis}
Many aspects of code, such as names, formatting, the lexical order of
methods, \etc have no impact on program semantics. This is precisely why we abstract them in most program analyses.
But then, why should statistical properties of code matter at all? To explain this, we recently suggested a hypothesis,
called the \emph{naturalness hypothesis}.
The inspiration for the naturalness hypothesis can be traced back to the ``literate programming'' 
concept of D. Knuth, which draws from the insight 
that programming is a form of human communication: 
\emph{``Let us change our traditional attitude to the construction of programs: Instead of imagining that our main task is to instruct a computer what to do, let us concentrate rather on explaining to human beings what we want a computer to do...'' ~\cite{knuth1984literate} }
The naturalness hypothesis, then, holds that 
\begin{quote}
\textbf{The naturalness hypothesis.} \emph{ Software is a form of human communication; software corpora have similar
statistical properties to natural language corpora;  and these properties can be exploited
to build better software engineering tools}. 
\end{quote}

The exploitation of the statistics of human communication is a mature and effective technology,
with numerous applications~\citep{jurafsky2000speech}.
Large corpora of human communication, \viz natural language corpora, have been
extensively studied, and highly refined statistical models of these corpora have been used
to great effect in speech recognition, translation, error-correction, \etc.

The naturalness hypothesis holds that, because
coding is an act of communication,  one might expect large code corpora to have rich patterns, similar 
to natural language, thus allowing software engineering tools to exploit probabilistic ML models. 
The first empirical evidence of this hypothesis, showing that models originally developed for 
natural language were surprisingly effective for source code, was
presented by \citet{hindle2012naturalness, hindle2016naturalness}. 
More evidence and numerous applications of this approach have followed, which are the subject
of this review.

The naturalness hypothesis, then, inspires the goal
to apply machine learning approaches to
create probabilistic source code models
that \emph{learn} how developers \emph{naturally} write and use code. These models can be used
to augment existing tools with statistical information and
enable new machine learning-based software engineering tools, such as recommender
systems and program analyses. At a high level, statistical methods allow a system to make hypotheses,
along with probabilistic confidence values, of what a developer \emph{might} want to do next
or what formal properties \emph{might} be true of a chunk of code. Probabilistic methods also provide
natural ways of learning correspondences between code and other types of documents, such as requirements,
blog posts, comments, \etc --- such correspondences will always be uncertain, because natural language
is ambiguous, and so the quantitative measure of confidence provided by probabilities is especially natural.
As we discuss in \autoref{sec:applications}, one could go so far as to claim that almost every area of software engineering and programming
language research has potential opportunities for exploiting statistical properties.

Although the ``naturalness hypothesis'' may not seem surprising,
one should appreciate the root cause of ``naturalness''. Naturalness of code seems
to have a strong connection with the fact that developers  prefer to write \citep{allamanis2014learning}
and read \citep{hellendoorn2015will} code that is conventional, idiomatic, and familiar
because it helps
understanding and maintaining software systems. Code that takes familiar forms
is more \emph{transparent}, in that its meaning is more readily apparent to an experienced reader. 
Thus, the naturalness hypothesis leads seamlessly to a  
``code predictability'' notion, suggesting that
code artifacts --- from simple token sequences to formal verification statements --- contain
useful recurring and predictable patterns that can be exploited.
``Naturalness'' and ``big code'' should be viewed
as instances of a more general concept that there is exploitable regularity
across human-written code that can be ``absorbed'' and generalized by a learning component
that can transfer its knowledge and probabilistically reason about new code.

This article reviews the emerging area of machine learning and statistical natural language processing
methods applied to source code.
We focus on probabilistic models of code, that is, methods that estimate a distribution
over all possible source files. Machine learning in probabilistic models has seen wide application
throughout artificial intelligence, including natural language processing, robotics, and computer vision,
because of its ability to handle uncertainty and to learn in the face of noisy data.
One might reasonably ask why it is necessary to handle uncertainty and noise in software development
tools, when in many cases the program to be analyzed is known (there is no uncertainty about what
the programmer has written) and is deterministic.
In fact, there are several interesting motivations for incorporating probabilistic modeling into machine learning
methods for software development.
First, probabilistic methods offer a principled method for handling uncertainty and
fusing multiple, possibly ambiguous, sources of information.
Second, probabilistic models provide a natural framework
for connecting prior knowledge to data --- providing a natural framework
to 
design methods based on abstractions of statistical properties
of code corpora.
In particular, we often wish to infer relationships between source code and natural language text, such as comments, bug reports, requirements documents, documentation, search queries, and so on.
Because natural language text is ambiguous, it is useful
to quantify uncertainty in the correspondence between code and text.
Finally, when predicting program properties, probabilities provide a way to relax strict requirements on  soundness: we can seek unsound methods
that predict program properties based on statistical patterns
in the code, using probabilities as a way to quantify the
method's confidence in its predictions.

\section{Text, Code and Machine Learning}
\label{sec:codevsnl}

Programming languages narrow the gap between computers and the human mind:
they construct palatable abstractions out of a multitude of minute state
transitions. Source code has two audiences and is inherently \emph{bimodal}: it
communicates along two channels: one with humans, and one with computers. 
Humans must understand code to read and write it; computers must be
able to execute it. 
The bimodality of code drives the similarities and differences
between it and text. Below, we
discuss these similarities and differences with forward pointers to how they have
been exploited, handled, or remain open.  Although code and text are similar, code written in a
general-purpose programming languages, is a relatively new problem domain
for existing ML and NLP techniques.  \citet{hindle2016naturalness} not only showed that
exploitable similarity exists between the two via an \ngram language model,
but that code is even less surprising than text. 
Although it may seem manifestly obvious that code
and text have many differences, it is useful
to enumerate these differences carefully,
as this allows us to gain insight into
when techniques from NLP need to be modified 
to deal with code. 
Perhaps the most obvious difference
is that code is executable and has formal
syntax and semantics.
We close by discussing how source code
bimodality manifests itself as synchronization points between the algorithmic
and explanatory channels.

\paragraph{Executability}
All code is executable; text often is not.
So code is often semantically brittle --- small
changes (\eg swapping function arguments) can drastically change the meaning of
code; whereas natural language is more 
robust in that readers can often understand text even if it contains
mistakes.
Despite
the bimodal nature of code and its human-oriented modality, the
sensitivity of code semantics to ``noise'' necessitates the combination of
probabilistic and formal methods. For example, existing work builds
probabilistic models then applies strict formal constraints to filter their
output (\autoref{subsec:generativemdls}) or uses them to guide formal methods
(\autoref{subsec:progAnalysis}). Nevertheless, further research on bridging
formal and probabilistic methods is needed (\autoref{subsec:mlFd}).   

Whether it is possible to translate between
natural languages in a way that completely
preserves meaning is a matter of 
debate.  Programming languages, on the other hand, can be translated between
each other exactly, as all mainstream programming
languages are Turing-complete.  (That said, porting
real-world programs to new languages and platforms
remains challenging in practice~\citep{brown1979software}).  ML techniques have not yet 
comprehensively tackled such problems and are currently limited to solely translating
among languages with very similar characteristics, 
\eg Java and C\# (\autoref{subsec:mlFd}).  Programming languages differ in their
expressivity and intelligibility, ranging from Haskell to
Malbolge\footnote{\url{https://en.wikipedia.org/wiki/Malbolge}}, with some
especially tailored for certain problem domains;  in contrast, natural
languages are typically used to communicate
across a wide variety of domains.  Executability of code induces control and data
flows within programs, which have only
weak analogs in text.  Finally, executability gives
rise to additional modalities of code --- its static and dynamic views (\eg
execution traces), which are not present in text.  Learning over traces or
flows are promising directions (\autoref{subsec:mlFd}).

\paragraph{Formality}

Programming languages are formal languages, whereas
formal languages are only mathematical models
of natural language.
As a consequence, programming languages are designed
top-down by a few designers for many users.  Natural languages, in contrast,
emerge, bottom up, ``through social dynamics''~\citep{croft2008evolutionary}.
Natural languages change gradually, while programming languages exhibit
punctuated change:  new releases, like Python 3, sometimes break backward
compatibility.  Formatting can also be
meaningful in code: Python's whitespace sensitivity is the canonical example.
Text has a robust environmental dependence, whereas code suffers from bit rot
--- the deterioration of software's functionality through time because of changes
in its environment (\eg dependencies) --- because all its
explicit environmental interactions must be specified upfront and execution
environments evolve much more quickly than natural languages.  

Source code's formality facilities reuse.  Solving a problem algorithmically is cognitively
expensive, so developers actively try to reuse code \citep{hunt2000pragmatic},
moving common functionality into libraries to facilitate reuse.  
As a result, usually functions are semantically unique within a project.
Coding competitions or undergraduate projects are obvious exceptions.  In contrast, one can find
thousands of news articles describing an important global event. 
On the other
hand, \citet{gabel2010study} have found that 
locally, code is more pattern dense than text
(\autoref{subsec:generativemdls}). This has led to important
performance improvements on some applications, such as code completion
(\autoref{subsec:recomSystems}).

Because programming languages are automatically translated into machine code,
they must be syntactically, even to a first approximation semantically,
unambiguous\footnote{Exceptions exist, like
\url{http://po-ru.com/diary/ruby-parsing-ambiguities/} and
\url{https://stackoverflow.com/questions/38449606/ambiguity-in-multiple-inheritance},
but these rarely matter in practice.}.  In contrast to NLP models, which must
always account for textual ambiguity, probabilistic models of code can and do
take advantage of the rich and unambiguous code structure.  Although it is less
pervasive, ambiguity remains a problem in the
analysis of code, because of issues like
polymorphism and aliasing.  \autoref{subsec:bg:representationalmdls} and
\autoref{subsec:migration} discuss particularly notable approaches to
handling them.  Co-reference ambiguities can arise when viewing code statically,
especially in dynamically typed languages (\autoref{subsec:mlFd}). The undefined
behavior that some programming languages permit can cause semantic ambiguity
and, in the field, syntactic problems can arise due to nonstandard
compilers~\citep{bessey2010few}; however, the dominance of a handful of
compilers/interpreters for most languages 
ameliorates both problems.

\paragraph{Cross-Channel Interaction} Code's two channels, the algorithmic
and the explanatory channels, interact through
their semantic units, but mapping code units to textual units remains an open
problem.  Natural semantic units in code are identifiers,
statements, blocks, and functions.  None of these universally maps to textual
semantic units.  For example, identifiers, even verbose function names that seek to describe
their function, carry less information than words like ``Christmas'' or ``set''.
In general, statements in code and sentences in text differ in how much background
knowledge the reader needs in order to understand them in isolation;
an arbitrary statement is far more likely to use domain-specific,
even project-specific, names or neologisms than an arbitrary sentence is.
Blocks vary greatly in length and semantics richness and often
lack clear boundaries to a human reader.  Functions are clearly delimited and
semantically rich, but long.  In text, a sentence is the natural multiword
semantic unit and usually contains fewer than 50 words (tokens).
Unfortunately, one cannot, however, easily equate them.  A function differs
from a sentence or a sequence of sentences, \ie a paragraph, in that it is
named and called, while, in general settings, sentences or paragraphs rarely 
have names or are referred to elsewhere in a text.
Further, a single function acts on the world,
so it is like a single \emph{action} sentence, but is usually much longer, 
 often containing hundreds of tokens, and usually performs multiple actions,
making a function closer to a sequence of sentences, or a paragraph, but 
paragraphs are rarely solely composed of action sentences.

Additionally, parse trees of sentences in text tend to be diverse, short, and shallow
compared to abstract syntax trees of functions, which are usually much deeper with
repetitive internal structure.  Code bases are multilingual (\ie contain code
in more than one programming language, \eg Java and SQL) with different tasks
described in different languages, more frequently than text corpora; this can
drastically change the shape and frequency of its semantic units.  Code has a
higher neologism rate than text.  Almost 70\% of all characters are identifiers
and a developer must choose a name for each
one~\citep{deissenboeck2006concise}; when writing text, an author rarely names
new things but usually chooses an existing word to use.  Existing work handles 
code's neologism rate by introducing cache mechanisms or decomposing
identifiers at a subtoken level (\autoref{subsec:generativemdls}).

Determining which semantic code unit is most useful for which task is an open
question.  
Consider the problem of automatically generating
comments that describe code, which can be
formalized as a machine translation problem
from code to text.
  Statistical machine
translation approaches learn from an aligned corpus.  Statement-granular
alignment yields redundant comments, while function granular alignment has
saliency issues (\autoref{sec:applications:SourceCodeAndNaturalLanguage}).
As another example, consider code search, where search engines must map 
queries into semantic code units.
  Perhaps the answer 
will be in maps from code to text whose units 
vary by granularity or context (\autoref{subsec:mlFd}).

\section{Probabilistic Models of Code}
\label{sec:probmodels}

In this section, we turn our attention to probabilistic machine learning models of source code.
A probabilistic model of source code is a probability distribution 
over code artifacts.
As all models do, probabilistic machine learning models 
make simplifying assumptions about the modeled domain. These
assumptions make the models tractable to learn and use, but introduce
error. Since each model makes different 
assumptions, each model has its own strengths and weaknesses and is
more suitable for some applications. In this section, we 
group existing work into families of models and discuss their assumptions.
To group these family of models
in terms of shared design choices,
we separate these models into three categories,
based on the form of the equation of the modeled probability 
distribution and their inputs and outputs,
with the caveat that some models fall into multiple categories. We 
also discuss how and why these models differ from their natural language counterparts.
In \autoref{sec:applications}, we discuss 
applications of these models in software engineering and programming
languages.

\begin{description}

  \item[Code-generating Models] define
  a probability distribution over code by
    stochastically modeling the generation of smaller and simpler parts of code, \eg
    tokens or AST nodes.

  \item[Representational Models of Code] take an abstract representation\footnote{
    In the machine learning literature, \emph{representation}, applied to code, is
    roughly equivalent to \emph{abstraction} in programming language research:
    a lossy encoding that preserves a semantic property of interest.} of
    code as input.  Example representations include token contexts or data flow.
    The resulting model yields a conditional probability distribution over code
    element properties, like the types of variables, and can predict them.

  \item[Pattern Mining Models] infer, without supervision, a likely latent
    structure within code. These models are an instantiation of clustering
    in the code domain; they can find reusable and human-interpretable patterns.

\end{description}

Code-generating models find analogues in
generative models of text, such as language models and machine translation models.
Code representational models are analogous to systems for named entity
recognition, text classification, and sentiment analysis in NLP. Finally, code pattern mining 
models are analogous to probabilistic topic models and ML techniques for mining
structured information (\eg knowledge-bases) from text.
To simplify notation below, we use $\codevec{c}$ to denote
an arbitrary code abstraction, like an AST.

\subsection{Code-generating Probabilistic Models of Source Code}
\label{subsec:generativemdls}
\newcommand{\codecx}{\mathcal{C}(\codevec{c})}
\newcommand{\data}{\mathcal{D}}

Code-generating probabilistic models of code are probability distributions
that describe a stochastic process for generating valid code, \ie they model
\emph{how} code is written. Given training data $\data$,
an output code representation $\codevec{c}$,
and a possibly empty context $\codecx$, these models
learn the probability distribution $P_\data(\codevec{c}|\codecx)$ and sample 
$P_\data$ to generate code.
When $\codecx=\varnothing$, the probability distribution $P_\data$ is a \emph{language model}
of code, \ie it models how code is generated when no external context information
is available. When $\codecx$ is a non-code modality (\eg natural language),
$P_\data$ describes a \emph{code-generative multimodal
model} of code. When $\codecx$ is also code, the probability
distribution $P_\data$ is a \emph{transducer model} of code. In addition to 
generating code, by definition, code generating probabilistic models act as a \emph{scoring function},
assigning a non-zero probability to every possible snippet of code. This score,
sometimes referred to as ``naturalness'' of the code~\citep{hindle2016naturalness}, suggests how probable
the code is under a learned model.

Since code-generating models predict the complex structure of code,
they make simplifying assumptions about the generative process
and iteratively predict elements of the code to generate a full
code unit, \eg code file or method.
Because of code's structural complexity and the simplifying assumptions
these models make to cope with it, none of the existing models in the literature
generate code that \emph{always} parses, compiles, and
executes. Some of the models do, however, impose constraints that take code
structure into account to remove some inconsistencies; 
for instance, \citep{maddison2014structured} only generate variables declared within each scope.

We structure the discussion about code-generating models of code as follows. We
first discuss how models in this category generate code and then we show how
the three different types of models (language, transducer and multimodal models)
differ.

\subsubsection{Representing Code in Code-Generating Models}
Probabilistic models for generating structured objects are widely in use in machine learning
and natural language processing with a wide range of applications. Machine learning
research is considering a wide range of structures from natural language sentences to
chemical structures and images. Within the source code domain, we can broadly
find three categories of models based on the way they generate code's structure:
token-level models that generate code as a sequence of tokens, syntactic
models that generate code as a tree and semantic models that generate
graph structures. Note that this distinction is about the generative process
and \emph{not} about the information used within this process. For example \citet{nguyen2013statistical}
uses syntactic context, but is classified as a token-level model that generates tokens.

\paragraph{Token-level Models (sequences)}
Sequence-based models are commonly used because of their simplicity. They view code
as a sequence of elements, usually code tokens or characters, \ie $\codevec{c}=t_{1}\dots t_{M}.$ 
Predicting a large sequence in a single step is infeasible due to the
exponential number of possible sequences; for a set of $V$ elements, there
are ${|V|}^N$ sequences of length $N$. Therefore, most sequence-based models
predict sequences by sequentially generating each element, \ie they model
the probability distribution $P(t_m|t_1 \dots t_{m-1}, \codecx)$. However, directly modeling
this distribution is impractical and all models make
different simplifying assumptions.

The \ngram model has been a widely used sequence-based model, most
commonly used as a language model. 
It is an effective and practical LM
for capturing local and simple statistical dependencies in sequences.
\ngram models assume that tokens are generated sequentially, left-to-right and 
that the next token can be
predicted using only the previous $n-1$ tokens.
The consequence of capturing a short context is that \ngram models
cannot handle long-range dependencies, notably scoping information. Formally,
the probability of a token $t_{m}$, is conditioned on the context $\codecx$ (if any)
and the generated sequence so far
$t_{1}\dots t_{m-1}$, which is assumed to depend on only the previous $n-1$ tokens.
Under this assumption, we write
\begin{equation}
\label{eq:ngramLM}
P_\data(\codevec{c}| \codecx)=P(t_{1}\dots t_{M} | \codecx)=\prod_{m=1}^{M}P(t_m | t_{m-1}\dots t_{m-n+1}, \codecx).
\end{equation}
To use this equation, we need to know the conditional probabilities
$P(t_m | t_{m-1}\dots t_{m-n+1}, \codecx)$ for each possible \ngram and context.
This is a table of $|V|^n$ numbers for each context $\codecx$.
These are the \emph{parameters} of the model
that we learn from the {training corpus}.
The simplest way to estimate the model parameters is to
set $P(t_m | t_{m-1}\dots t_{m-n+1})$
to the proportion of times that $t_m$ follows $t_{m-1}\dots t_{m-n+1}$.
In practice, this simple estimator does not work well, because it assigns
zero probability to
\ngrams that do not occur in the training corpus.
Instead, \ngram models use \emph{smoothing} methods
\citep{chen1999empirical} as a principled way for assigning
probability to unseen \ngrams by extrapolating information from $m$-grams ($m<n$).
Furthermore, considering \ngram models with non-empty contexts $\codecx$
exacerbates sparsity rendering these models impractical.
Because of this, \ngrams are predominantly used as language models.
The use of \ngram LMs in software engineering originated with the
pioneering work of \citet{hindle2012naturalness} who used an \ngram
LM with Kneser-Ney \citep{kneser1995improved} smoothing. Most subsequent
research has followed this practice.

In contrast to text, code tends to be more verbose \citep{hindle2012naturalness}
and much information is lost within the $n-1$ tokens of the context. To
tackle this problem, \citet{nguyen2013statistical} extended
the standard \ngram model by annotating the code tokens with parse information that can
be extracted from the currently generated sequence. This increases the available
context information allowing the \ngram model to achieve better predictive performance.
Following this trend, but using concrete and abstract semantics of code,
\citet{raychev2014code} create a token-level model that treats code generation
as a combined synthesis and probabilistic modeling task. 

\citet{tu2014localness} and later, 
\citet{hellendoorn2017deep} noticed that code has a high degree of \emph{
localness}, where identifiers (\eg variable names) are repeated often within close
distance. In their work, they adapted work in speech and natural language processing \citep{kuhn1990cache}
adding a cache mechanism that assigns higher probability to tokens that have been
observed most recently, achieving significantly better performance compared to other \ngram
models. Modeling identifiers in code is challenging~\citep{allamanis2013mining,maddison2014structured,allamanis2015suggesting,bielik2016phog}.
The agglutinations of multiple subtokens (\eg in \id{getFinalResults})
when creating identifiers is one reason. Following recent NLP work that models subword structure
(\eg morphology)~\citep{sennrich2065neural}, explicitly modeling subtoken in
identifiers may improve the performance of generative models.
Existing token-level code-generating models do not produce syntactically valid
code. 
\citet{raychev2014code} added additional context in the form of constraints --- derived from program analysis ---
to avoid generating some incorrect code.

More recently, sequence-based code models have turned to deep recurrent neural network (RNN)
models to outperform \ngrams. 
These models predict each token sequentially, but loosen the fixed-context-size
assumption, instead representing the context using a distributed
vector representation (\autoref{subsec:bg:representationalmdls}).
Following this trend, \citet{karpathy2015visualizing} and \citet{cummins2017synthesizing}
use character-level LSTMs \citep{hochreiter1997long}. Similarly, \citet{white2015toward} and \citet{dam2016deep}
use token-level RNNs. Recently, \citet{bhoopchand2016learning}
used a token sparse pointer-based neural model of Python that
learns to copy recently declared identifiers to capture very long-range dependencies
of identifiers, outperforming standard LSTM models\footnote{This work
differs from the rest from the fact that it anonymizes/normalizes identifiers, creating a less
sparse problem. Because of the anonymization, the results are not directly comparable
with other models.}.

Although neural models
usually have superior predictive performance, training them is significantly
more costly compared to \ngram models usually requiring orders of magnitude more data.
Intuitively, there are two reasons why deep learning
methods have proven successful for language models. First, the hidden state in an RNN
can encode longer-range dependencies of variable-length beyond the
short context of \ngram models.
Second, RNN language models can learn a much richer
notion of similarity across contexts.
For example, consider an $13$-gram model over code, in which we are trying to
estimate the distribution following the context
\lstinline+for(int i=N; i>=0; i--)+.
In a corpus, few examples
of this pattern may exist
because such long contexts occur rarely.
A simple \ngram model cannot exploit
the fact
that this context is very similar to 
\lstinline+for(int j=M; j>=0; j--)+.
But a neural network \emph{can} exploit it, by learning to assign
these two sequences similar
vectors.

\paragraph{Syntactic Models (trees)} Syntactic (or structural) code-generating models
 model code at the level of abstract syntax trees (ASTs). Thus, in contrast to
sequence-based models, they describe a stochastic process of generating tree structures. 
Such models make simplifying assumptions about how a tree is generated,
usually following generative NLP models of 
syntactic trees: 
they start from a root node, then sequentially generate children
top-to-bottom and left-to-right.  Syntactic models
generate a tree node conditioned on context defined as the forest of subtrees 
generated so far. In contrast to sequence models, these models --- by construction
--- generate syntactically correct code. In general, 
learning models that generate tree structures is harder compared to generating sequences:  it
is relatively computationally expensive, especially for neural models, given the variable shape and
size of the trees that inhibit efficient batching.
In contrast to their wide application in NLP,
probabilistic context free grammars 
(PCFG) have been found to be unsuitable as
language models of code  \citep{maddison2014structured,raychev2016learning}.
This may seem surprising, because most parsers assume that programming languages are context-free.
But the problem is that the PCFGs are not a good
model of \emph{statistical} dependencies
between code tokens, 
because nearby tokens may be far away in the AST. So it is not that PCFGs do not
capture long-range dependencies (\ngram-based models do not either), but that they do not
even capture close-range dependencies that matter~\cite{bielik2016phog}. Further, ASTs tend to be
deeper and wider than text parse trees due to the highly
compositional nature of code.

\citet{maddison2014structured} and \citet{allamanis2015bimodal} increase the size of the context considered
by creating a non-context-free log-bilinear neural network grammar, using
a distributed vector representation for the context. Additionally, \citet{maddison2014structured}
restricts the generation to generate variables that have been declared.
To achieve this, they use the deterministically-known information and
filter out invalid output. This simple process 
always produces correct code, even when the network does not learn to produce
it.  In contrast, \citet{amodio2017neural}
create a significantly more complex model that aims to learn to
enforce deterministic constraints of the code generation, rather than
enforcing them on the directly on the output. We further discuss the
issue of embedding constraints and problem structure in models \vs
learning the constraints in \autoref{subsec:futuredirs}.
 
\citet{raychev2016learning,bielik2016phog} increase the context by
annotating PCFGs with a learned program that uses features from the code.
Although the programs can, in principle, be arbitrary, they limit themselves to
synthesizing decision tree programs.
Similarly,
\citet{wang2016neural,yin2017syntactic} use an LSTM over AST nodes to achieve the same goal.
\citet{allamanis2014mining} also create a syntactic model learning 
Bayesian TSGs \citep{cohn2010inducing,post2009bayesian}
(see \autoref{subsec:unsupervisedmdls}).

\paragraph{Semantic Models (graphs)}
Semantic code-generating models view code as a graph. Graphs
are a natural representation of source code that require little abstraction
or projection. Therefore, graph model can be thought as generalizations of sequence and tree
models. However, generating complex graphs is hard, since there
is no natural ``starting'' point or generative process, as reflected by the limited number
of graphs models
in the literature. We refer the interested reader to the related work section
of \citet{johnson2016learning} for a discussion of recent models in the machine learning literature. To our knowledge, there are no
generative models that directly generate graph representations of realistic code (\eg data-flow
graphs). 
\citet{nguyen2015graph} propose a generative model, related to graph generative models in
NLP, that suggests application programming interface (API) completions.  They train their model over API
usages. However, they predict entire graphs as completions and
perform no smoothing, so their model will assign zero probability to unseen
graphs. In this way, their model differs from graph generating models in NLP, which
can generate arbitrary graphs.

\subsubsection{Types of Code Generating Models}
\label{sec:mlbackground:gentypes}
We use external context $\codecx$ to refine 
code generating models into three subcategories.

\paragraph{Language Models}
Language models
model the language itself, without using any external context, \ie $\codecx=\varnothing$.
Although LMs learn the
high-level structure and constraints of programming
languages fairly easily, predicting and generating source code identifiers (\eg
variable and method names), long-range dependencies and taking into account
code semantics makes the language modeling of code a hard and interesting 
research area. We discuss these and other differences and their implications
for probabilistic modeling in \autoref{subsec:futuredirs}.

Code LMs are evaluated like LMs in NLP,
using perplexity (or equivalently cross-entropy) and word error rate. Cross-entropy $H$ is the 
most common measure. Language models --- as most predictive machine learning
models --- can be seen as compression algorithms where the model 
predicts the full output (\ie decompresses) using extra information.
Cross-entropy measures
the average number of extra bits of information per token of code
that a model needs to decompress the correct output using a perfect
code (in the information-theoretic sense)
\begin{equation}
 H(\codevec{c}, P_\data) = -\frac{1}{M}\log_2 P_\data(\codevec{c})
\end{equation}
where $M$ is the number of tokens within $\codevec{c}$. By convention, the average
is reported per-token, even for non-token models. Thus, a ``perfect'' model, correctly predicting
all tokens with probability 1, would require no additional bits of information because,
in a sense, already ``knows'' everything. 
Cross-entropy allows comparisons across different models. 
Other, application-specific measures, are used  
when the LM was trained for a specific task, such as code completion (\autoref{subsec:recomSystems}). 

\paragraph{Code Transducer Models}
Inspired by
statistical machine translation (SMT), transducer models translate/transduce code from
one format into another (\ie $\codecx$ is also code), such as translating code from one source language into another,
target language. They have the form $P_\data(\codevec{c}|\codevec{s})$, where
$\codevec{c}$ is the target source code that is generated and $\codecx=\codevec{s}$ is the
source source code.
Most code transducer models use phrase-based machine translation.
Intuitively, phrase-based
 models assume that small chunks from the source input can
directly be mapped to chunks in the output. Although this assumption is
reasonable in NLP and many source code tasks, these models present challenges
in capturing long-range dependencies within the source and target. For example,
as we will mention in the next section, transducing code from an
imperative source language to a functional target is not currently possible
because of the source and target are related with a significantly more
complicated relation that simply matching ``chunks'' of the input code to
``chunks'' in the output.

These types of models have found application within code migration \citep{aggarwal2015using,nguyen2015divide,karaivanov2014phrase},
pseudocode generation \citep{oda2015learning} and code fixing \citep{pu2016skp}.
Traditionally transducer models have followed a
 noisy channel model, in which they combine a language model
$P_\data(\codevec{c})$ of the target language with a translation/transduction model $P_\data(\codevec{s}|\codevec{c})$ to match
elements between the source and the target. These methods pick the optimal transduction $c^*$ such that
$\codevec{c}^* = \arg\max P_\data(\codevec{c}|\codevec{s}) = \arg\max P_\data(\codevec{s}|\codevec{c}) P_\data(\codevec{c}),$
where the  second equality derives from the Bayes equation.
Again, these probabilistic generative models of code do
\emph{not} necessarily produce valid code, due to the simplifying
assumptions they make in both $P_\data(\codevec{c})$ and 
$P_\data(\codevec{s}|\codevec{c})$.
More recently, machine translation methods
based on phrase-based models and the noisy channel
model have been outperformed by neural network-based methods
that directly model $P_\data(\codevec{c}|\codevec{s})$. 

 Transducer models can
be evaluated with SMT evaluation measures, such as BLEU \citep{papineni2002bleu}
--- commonly used in machine translation as an approximate measure of translation quality ---
or programming and logic-related measures (\eg \emph{``Does the translated code parse/compile?''} and
\emph{``Are the two snippets equivalent?''}).

\paragraph{Multimodal Models}
Code-generating multimodal models correlate code
with one or more
non-code modalities, such as comments,
specifications, or search queries. These models have the form $P_\data(\codevec{c}|\vect{m})$
\ie $\codecx=\vect{m}$ is a representation of one or more non-code modalities. Multimodal
models are closely related to representational models (discussed
in \autoref{subsec:bg:representationalmdls}): multimodal code-generating models
learn an intermediate representation of the non-code modalities $\vect{m}$ and
use it to generate code. In contrast, code representational models create
an intermediate representation of the code but are \emph{not} concerned with
code generation. 

Multimodal models of code have been used for code synthesis, where 
the non-code modalities are leveraged to estimated a conditional generative
model of code, \eg synthesis of code given a natural language description by
\citet{gulwani2014nlyze} and more recently by \citet{yin2017syntactic}.
The latter model is a syntactic model that accepts natural language.
Recently, \citet{beltramelli2017pix2code,ellis2017learning,deng2017image} designed multimodal model
that accept visual input (the non-code modality) and generate code in a DSL describing 
how the input (hand-drawn image, GUI screenshot) was constructed.
Another use of these models is to score the co-appearance of the modalities,
\eg in code search, to score the probability of some text given a textual
query~\citep{allamanis2015bimodal}. This stream of research is related to
work in NLP and computer vision where one seeks to generate
a natural language description for an image.
These models are closely related to the other code generating models,
since they generate code. 
These models make also assume that the input modality conditions the
generation process. Multimodal models combine an assumption with a design choice. 
Like language models, these models
assume that probabilistic models can capture the process by which developers generate code; 
unlike language models, they additionally bias code generation using information from the
input modality $\vect{m}$.
The design choice is how to transform
the input modality into an intermediate representation. For example, \citet{allamanis2015bimodal}
use a bag-of-words assumption averaging the words' distributed representations.
However, this limits the expressivity of the models because the input modality
has to fit in whole within the distributed representation. To address this issue,
\citet{ling2016latent} and \citet{yin2017syntactic} use neural attention mechanisms
to selectively attend to information within the input modality without the
need to ``squash'' all the information into a single representation. 
Finally, the text-to-code problem, in
which we take the input modality $\vect{m}$ 
to be natural language text and the
other modality $\codevec{c}$ to be code,
 is 
 closely related to the problem of semantic parsing in NLP; see \autoref{sec:applications:SourceCodeAndNaturalLanguage}.

\newcommand{\Language}{$P_\data(\codevec{c})$}
\newcommand{\Translation}{$P_\data(\codevec{c}|\codevec{s})$}
\newcommand{\Multimodal}{$P_\data(\codevec{c}|\vect{m})$}
\begin{table}[tbp]
  \caption{Research on Source Code Generating Models $P_\data(\codevec{c}|\codecx)$ (sorted alphabetically),
  describing the process of generating source code.
 References annotated with $^*$ are also included in other categories.}\label{tbl:generativemodels}
 \begin{minipage}{\columnwidth}\begin{center}\footnotesize
  \begin{tabular}{llp{2.3cm}p{2.75cm}p{3cm}}\toprule
        Reference & Type & Representation & $P_\data$ & Application \\ \midrule
        \citet{aggarwal2015using} & \Translation & Token & Phrase & Migration \\
        \citet{allamanis2013mining} & \Language & Token & \ngram & --- \\
        \citet{allamanis2014learning}& \Language & Token + Location & \ngram & Coding Conventions \\
        \citet{allamanis2014mining}$^*$ & \Language & Syntax & Grammar (pTSG) & ---\\
        \citet{allamanis2015bimodal}$^*$ & \Multimodal &  Syntax & Grammar (NN-LBL)& Code Search/Synthesis \\
        \citet{amodio2017neural} & \Language & Syntax$\,+\,$Constraints & RNN & ---\\
        \citet{barone2017parallel} & \Multimodal & Token & Neural SMT & Documentation \\
        \citet{beltramelli2017pix2code} & \Multimodal & Token & NN (Encoder-Decoder) & GUI Code Synthesis\\
        \citet{bhatia2016automated} & \Language & Token & RNN (LSTM) & Syntax Error Correction \\
        \citet{bhoopchand2016learning} & \Language & Token & NN (Pointer Net) & Code Completion\\
        \citet{bielik2016phog} & \Language & Syntax & PCFG + annotations & Code Completion \\
        \citet{campbell2014syntax} & \Language & Token & \ngram & Syntax Error Detection \\
        \citet{cerulo2015irish} & \Language & Token & Graphical Model (HMM) & Information Extraction\\
        \citet{cummins2017synthesizing} & \Language & Character & RNN (LSTM) & Benchmark Synthesis\\
        \citet{dam2016deep} & \Language & Token & RNN (LSTM)& ---\\
        \citet{gulwani2014nlyze} & \Multimodal & Syntax & Phrase Model & Text-to-Code \\
        \citet{gvero2015synthesizing} & \Language & Syntax & PCFG + Search & Code Synthesis\\
        \citet{hellendoorn2015will} & \Language & Token & \ngram & Code Review\\
        \citet{hellendoorn2017deep} & \Language & token & \ngram (cache) & -- \\
        \citet{hindle2012naturalness} & \Language & Token & \ngram & Code Completion \\
        \citet{hsiao2014using} & \Language & PDG & \ngram & Program Analysis \\
        \citet{lin2017program} & \Multimodal & Tokens &  NN (Seq2seq) & Synthesis \\
        \citet{ling2016latent} & \Multimodal & Token & RNN + Attention & Code Synthesis \\
        \citet{liu2016towards} & \Language & Token & \ngram & Obfuscation\\
        \citet{karaivanov2014phrase} & \Translation & Token & Phrase & Migration\\
        \citet{karpathy2015visualizing} & \Language & Characters & RNN (LSTM)& --- \\
        \citet{kushman2013using} & \Multimodal & Token & Grammar (CCG) & Code Synthesis\\
        \citet{maddison2014structured} & \Language & Syntax with scope & NN & --- \\
        \citet{menon2013machine} & \Multimodal & Syntax & PCFG + annotations & Code Synthesis\\
        \citet{nguyen2013lexical} & \Translation & Token & Phrase & Migration \\
        \citet{nguyen2013statistical} & \Language & Token + parse info & \ngram & Code Completion \\
        \citet{nguyen2015divide} & \Translation & Token + parse info & Phrase SMT & Migration \\
        \citet{nguyen2015graph} & \Language & Partial PDG &  \ngram & Code Completion \\
        \citet{oda2015learning} & \Translation & Syntax + Token & Tree-to-String + Phrase & Pseudocode Generation\\
        \citet{patra2016learning} & \Language & Syntax & Annotated PCFG & Fuzz Testing\\
        \citet{pham2016learning} & \Language & Bytecode & Graphical Model (HMM) & Code Completion \\
        \citet{pu2016skp} & \Translation & Token & NN (Seq2seq) & Code Fixing\\
        \citet{rabinovich2017abstract}$^*$ & \Multimodal & Syntax & NN (LSTM-based)& Code Synthesis\\
        \citet{raychev2014code} & \Language & Token + Constraints & \ngram/ RNN & Code Completion \\
        \citet{ray2015naturalness} & \Language & Token & \ngram (cache) & Bug Detection \\
        \citet{raychev2016learning} & \Language & Syntax & PCFG + annotations & Code Completion \\
        \citet{saraiva2015products} & \Language & Token & \ngram & --- \\
        \citet{sharma2015nirmal} & \Language & Token & \ngram & Information Extraction\\
        \citet{tu2014localness} & \Language & Token & \ngram (cache) & Code Completion \\
        \citet{vasilescu2017recovering} & \Translation & Token & Phrase SMT & Deobfuscation\\
        \citet{wang2016neural} & \Language & Syntax & NN (LSTM)& Code Completion\\
        \citet{white2015toward} & \Language & Token & NN (RNN) & --- \\
        \citet{yadid2016extracting} & \Language & Token & \ngram & Information Extraction\\ 
        \citet{yin2017syntactic} & \Multimodal & Syntax & NN (Seq2seq) & Synthesis \\\bottomrule
    \end{tabular}
    {\scriptsize Abbreviations:
    \begin{inparaitem} 
      \item pTSG: probabilistic tree substitution grammar
      \item NN: neural network
      \item LBL: log-bilinear
      \item SMT: statitstical machine translation
      \item PCFG: probabilistic context-free grammar
      \item HMM: hidden Markov model
      \item LSTM: long short-term memory
      \item RNN: recurrent neural network
      \item CCG: combinatory categorial grammar
      \item Seq2seq: sequence-to-sequence neural network
    \end{inparaitem}}
  \end{center}
\end{minipage}
\end{table}

\subsection{Representational Models of Source Code}
\label{subsec:bg:representationalmdls}

Generative models recapitulate the process of generating source code, but cannot
explicitly predict facts about the code that may be directly useful to
engineers or useful for other downstream tasks, such as static analyses.
To solve this problem, researchers have built models to learn intermediate, not necessarily human-interpretable,
encodings of code, like a vector embedding.  
These models predict the probability distribution
of properties of code snippets, like variable types.
We call them representational code models.
They learn the conditional
probability distribution of a code property $\pi$ as $P_\data(\pi|f(\codevec{c}))$, where $f$ is
a function that transforms the code $\codevec{c}$ into a
target representation and $\pi$ can be an arbitrary set of features or other (variable) structures.
These models use a diverse set of machine learning methods
and are often application-specific. \autoref{tbl:representationalmodels} lists
representational code model research. Below we discuss two types of models.
Note that they are \emph{not} mutually exclusive and models frequently combine distributed representations 
and structured prediction.

\subsubsection{Distributed Representations}
\label{subsec:distributedReps}

Distributed representations \citep{hinton1984distributed} are widely used in
NLP to encode natural language elements. For example, \citet{mikolov2013efficient}
learn distributed representations of words, showing that such representations can learn
useful semantic relationships and \citet{le2014distributed} extend this idea to
sentences and documents.
Distributed representations refer to arithmetic vectors or matrices where the
meaning of an element is \emph{distributed} across multiple components (\eg the
``meaning'' of a vector is distributed in its components).
This contrasts with local representations,
where each element is uniquely represented with exactly one component. Distributed
representations are commonly used in machine learning and NLP because they 
tend to generalize better and have recently become extremely common due to
their omnipresence in deep learning.
Models that learn distributed representations assume that
the elements being represented and their relations can be encoded within a 
multidimensional real-valued space and that the relation (\eg similarity)
between two representations can be measured within this space.
Probabilistic code models widely use distributed representations. For example, models
that use distributed vector representations learn a function of the form
$\codevec{c} \rightarrow \R^D$ that maps code elements to a $D$-dimensional
vector. Such representations are usually the (learned) inputs or output
of (deep) neural networks. 

\citet{allamanis2015suggesting} learn distributed vector representations
for variable and methods usage contexts and use them to predict a probability 
distribution over their names. Such distributed representations are quite
similar to those produced by word2vec \citep{mikolov2013efficient};
the authors found that the distributed vector representations of variables and methods
learn common semantic properties, implying that some form
of the distributional hypothesis in NLP also holds for code.

\citet{gu2016deep} use a sequence-to-sequence deep neural network \citep{sutskever2014sequence}, originally introduced for
SMT, to learn intermediate distributed vector representations of 
natural language queries which
they use to predict relevant API sequences. \citet{mou2016convolutional}
learn distributed vector representations using custom convolutional neural networks
to represent features of 
snippets of code, then they assume that student solutions to various coursework problems have been intermixed and seek to recover the solution-to-problem mapping via classification.

\citet{li2015gated} learn
distributed vector representations for the nodes of a memory heap and use the
learned representations to synthesize candidate
formal specifications for the code that produced the heap.
\citet{li2015gated} exploit heap structure to define graph neural networks,
a new machine learning model based on gated recurrent units (GRU, a type of RNN \citep{cho2014properties}) to directly learn from heap
graphs.
\citet{piech2015learning} and \citet{parisotto2016neuro} learn distributed representations of source
code input/output pairs and use them to assess and review student assignments
or to guide program synthesis from examples.

Neural code-generative models of code also use distributed representations 
to capture context, a common practice 
in NLP.
For example, the work of \citet{maddison2014structured} and other neural language
models (\eg LSTMs in \citet{dam2016deep})
describe context distributed representations while sequentially
generating code. \citet{ling2016latent} and \citet{allamanis2015bimodal}
combine the code-context distributed representation with a distributed representations of
other modalities (\eg natural language) to synthesize code. While all 
of these representations can, in principle, encode unbounded context, 
handling all code dependencies of arbitrary length is an unsolved problem.
Some neural architectures, such as LSTMs \citep{hochreiter1997long}, GRUs \citep{cho2014properties}
and their variants, have made progress on this problem and can handle moderately long-range dependencies.

\subsubsection{Structured Prediction}
Structured prediction is the problem of predicting  a set of interdependent variables, 
given a vector of input features.
Essentially, structured prediction generalizes
standard classification to multiple output variables.
A simple example of structured prediction
is to predict a part-of-speech tag for each
word in a sentence.
Often the practitioner defines a dependency structure
among the outputs, \eg, via a graph, as 
part of the model definition.
 Structured prediction has been widely 
studied within machine learning and NLP, and 
 are omnipresent in code. 
  Indeed, structured prediction is particularly well-suited to code,
because it can exploit the semantic and syntactic
 structure of code to define the model.
Structured prediction is a general framework
to which deep learning methods have been applied.
For example, the celebrated
sequence-to-sequence (seq2seq) learning 
models~\cite{sutskever2014sequence,bahdanau2014neural} are general methods
for tackling the related structured prediction problem. In short, structured prediction 
and distributed representations are not mutually exclusive.

One of the most well-known applications of structured
prediction to source code is
\citet{raychev2015predicting}, who represent code as a variable dependency network,
represent each JavaScript variable as a single node,
and model their pairwise interactions as a 
conditional random field (CRF). They train the CRF to jointly predict the
types and names of all variables within a snippet of code. \citet{proksch2015intelligent}
use a directed graphical model to represent the context of an (incomplete)
usage of an object to suggest a method invocation (\viz constructor) autocompletion 
in Java.

Structured prediction, such as predicting a sequence of elements, can be combined with distributed representations.
For example, \citet{allamanis2016convolutional,allamanis2015suggesting}
use distributed representations to predict sequences
of identifier sub-tokens to build a single token 
and \citet{gu2016deep} predict the sequence of API calls.
\citet{li2015gated} learn distributed representations
for the nodes of a fixed heap graph by considering its structure and the
interdependencies among the nodes. \citet{kremenek2007factor} use a factor graph to
learn and enforce 
API protocols, like the resource usage specification of the POSIX file API, as do \citet{livshits09merlin} for information flow 
problems.
\citet{allamanis2018learning} predict the data flow graph of code by learning to
paste snippets of code into existing code and adapting the variables used.

\begin{table}[ptb]
  \caption{Research on Representational Models of Source Code $P_\data(\pi|f(\codevec{c}))$ (sorted alphabetically).
   References annotated with $^*$ are also included in other categories. GM refers
   to graphical models}\label{tbl:representationalmodels}
    \begin{minipage}{\columnwidth}\begin{center}\footnotesize
  \resizebox{\textwidth}{!}{    
   \begin{tabular}{lp{2cm}p{2.2cm}p{2cm}p{2.7cm}}
        \toprule
        Reference & Input Code Representation ($\codevec{c}$) & Target ($\pi$) & Intermediate Representation ($f$) & Application \\ \midrule
        \citet{allamanis2015suggesting} & Token Context & Identifier Name & Distributed & Naming \\
        \citet{allamanis2015bimodal}$^*$ & Natural Language & LM (Syntax) & Distributed & Code Search\\
        \citet{allamanis2016convolutional} & Tokens & Method Name & Distributed & Naming\\
        \citet{allamanis2018learning} & PDG & Variable Use Bugs & Distributed & Program Analysis\\
        \citet{bavishi2017context2name} & Token Context & Identifier Name & Distributed & Naming \\
        \citet{bichsel2016statistical} & Dependency Net & Identifier Name & CRF (GM) & Deobfuscation \\
        \citet{bruch2009learning} & Partial Object Use & Invoked Method & Localized & Code Completion\\
        \citet{chae2016automatically} & Data Flow Graph & Static Analysis & Localized & Program Analysis\\
        \citet{corley2015exploring} & Tokens & Feature Location & Distributed & Feature Location\\
        \citet{cummins2017end} & Tokens & Optimization Flags &Distributed & Optimization Heuristics\\
        \citet{dam2016deep}$^*$ & Token Context & LM (Tokens) & Distributed & ---\\
        \citet{gu2016deep} & Natural Language & API Calls & Distributed & API Search\\
        \citet{guo2017semantically} & Tokens & Traceability link & Distributed & Traceability\\
        \citet{gupta2017deepfix} & Tokens & Code Fix & Distributed & Code Fixing\\
        \citet{gupta2018deep} & Tokens & Code Fix & Distributed & Code Fixing\\
        \citet{hu2017codesum} & Linearized AST & Natural Language & Distributed & Summarization\\
        \citet{iyer2016summarizing} & Tokens & Natural Language & Distributed & Summarization\\
        \citet{jiang2017automatically} & Tokens (Diff) & Natural Language & Distributed & Commit Message\\
        \citet{koc2017learning} & Bytecode & False Positives & Distributed & Program Analysis\\
        \citet{kremenek2007factor} & Partial PDG & Ownership & Factor (GM) & Pointer Ownership\\
        \citet{levy2017learning} & Statements& Alignment & Distributed & Decompiling\\
        \citet{li2015gated} & Memory Heap & Separation Logic & Distributed & Verification\\
        \citet{loyola2017neural} & Tokens (Diff) & Natural Language & Distributed & Explain code changes\\
        \citet{maddison2014structured}$^*$& LM AST Context & LM (AST) & Distributed & ---\\
        \citet{mangal2015user} & Logic + Feedback & Prob. Analysis & MaxSAT & Program Analysis\\
        \citet{movshovitz2013natural} & Tokens & Code Comments & Directed GM & Comment Prediction\\
        \citet{mou2016convolutional} & Syntax & Classification & Distributed & Task Classification\\
        \citet{nguyen2016mapping} & API Calls & API Calls & Distributed & Migration\\
        \citet{omar2013structured} & Syntactic Context & Expressions & Directed GM & Code Completion\\
        \citet{oh2015learning} & Features & Analysis Params & Static Analysis & Program Analysis\\
        \citet{piech2015learning} & Syntax + State & Student Feedback & Distributed & Student Feedback\\
        \citet{pradel2017deep} & Syntax & Bug Detection & Distributed & Program Analysis \\
        \citet{proksch2015intelligent} & Inc. Object Usage & Object Usage & Directed GM & Code Completion\\
        \citet{rabinovich2017abstract}$^*$ & LM AST Context & LM (AST) & Distributed & Code Synthesis\\
        \citet{raychev2015predicting} & Dependency Net & Types + Names & CRF (GM) & Types + Names\\
        \citet{wang2016bugram} & Tokens & Defects & LM (\ngram) & Bug Detection\\
        \citet{white2015toward}$^*$ & Tokens & LM (Tokens) & Distributed & --- \\
        \citet{white2016deep}$^*$ & Token + AST & --- & Distributed & Clone Detection\\
        \citet{zaremba2014learning} & Characters & Execution Trace & Distributed & ---\\ \bottomrule
  \end{tabular} 
  }  
\end{center}\end{minipage}
\end{table}

\subsection{Pattern Mining Models of Source Code}
\label{subsec:unsupervisedmdls}
Pattern mining models aim to discover a finite set of human-interpretable patterns from source code,
without annotation or supervision, and present the mined patterns to software engineers. 
Broadly, these models cluster source code into a finite set of groups. 
Probabilistic pattern mining models of code infer the likely latent structure of a
probability distribution
\begin{align}
P_\data(f(\codevec{c}))=\sum_\vect{l} P_\data(g(\codevec{c})|\vect{l})P(\vect{l})
\end{align}
where $g$ is a deterministic function that
returns a (possibly partial, \eg API calls only) view of the code and
$\vect{l}$ represents a set of latent variables that the model introduces and aims
to infer. Applications of
such models are common in the mining software repositories community and
include documentation (\eg API patterns), summarization, and anomaly detection.
\autoref{tbl:unsupervisedmodels} lists this work. 
Unsupervised learning is one of the most challenging areas in machine learning.
This hardness stems from the need to automatically distinguish important patterns
in the code from spurious patterns that may appear to be significant because
of limited and noisy data. When designing unsupervised models, the core
assumption lies in the objective function being used and often we resort to
using a principle from statistics, information theory or a proxy supervised
task. Like all machine learning models, they require assumptions 
about how the data is represented. An important issue with unsupervised
methods is the hardness of evaluating the output, since the quality of the output
is rarely quantifiable.
A vast literature on \emph{non-probabilistic} methods 
exploits data mining methods, such as frequent pattern mining and
anomaly detection~\citep{witten2016data}.  
We do \emph{not} discuss these models here, since they are \emph{not}
probabilistic models of code. 
Classic probabilistic topic models~\citep{blei12topic}, which usually views code (or 
other software engineering artifacts) as a bag-of-words, have also been heavily
investigated. Since these models and their
strengths and
 limitations are well-understood, we omit them here.

\citet{allamanis2014mining} learn a tree substitution grammar (TSG) using
Bayesian nonparametrics, a technique originally devised for natural language
grammars. TSGs learn to group commonly co-appearing grammar productions (tree fragments).
Although TSGs have been used in NLP to improve parsing performance (which is
ambiguous in text), \citet{allamanis2014mining} observe that the inferred
fragments represent common code usage conventions and name them \emph{idioms}. 
Later, \citet{allamanis2016mining} extend this technique to mine semantic code
idioms by modifying the input code representation and adapting the inference method.

In a similar fashion, \citet{fowkes2015parameter}
learn the latent variables of a graphical model to infer common API
usage patterns. Their method automatically infers the most probable grouping
of API elements. This is in stark contrast to frequency-based methods \citep{xie2006mapo}
that suffer from finding frequent but not necessarily
``interesting'' patterns.
Finally, \citet{movshovitz2015kb} infer the latent variables of
a graphical model that models a software ontology.

As in NLP and machine learning in general, evaluating pattern mining models is hard, since the quality of the discovered
latent structure is subjective. Thus, researchers often resort to 
extrinsic, application-specific measures. For example, \citet{fowkes2017autofolding}
run a user study to directly assess the quality of their summarization method.

\begin{table}[tb]
   \caption{Research on Pattern Mining Probabilistic Models of Source Code (sorted alphabetically).
      These models have the general form $P_\data(g(\codevec{c}))$.
      References annotated with $^*$ are also included in other categories.}
      \label{tbl:unsupervisedmodels}
       \begin{minipage}{\columnwidth}\begin{center}\footnotesize
  \begin{tabular}{lp{2.7cm}p{2.2cm}p{2.8cm}}
        \toprule
        Reference & Code Representation ($\codevec{c}$) & Representation ($g$) & Application \\ \midrule
        \citet{allamanis2014mining}$^*$& Syntax & Graphical Model & Idiom Mining\\
        \citet{allamanis2016mining}& Abstracted AST & Graphical Model & Semantic Idiom Mining\\
        \citet{fowkes2015parameter}& API Call Sequences & Graphical Model & API Mining\\
        \citet{murali2017bayesian} & Sketch Synthesis & Graphical Model & Sketch Mining\\
        \citet{murali2017finding}& API Usage Errors & Graphical Model & Defect Prediction\\
        \citet{movshovitz2015kb} & Tokens & Graphical Model & Knowledge-Base Mining\\
        \citet{nguyen2017exploring} & API Usage & Distributed & API Mining\\
        \citet{fowkes2017autofolding} & Tokens & Graphical Model & Code Summarization\\
        \citet{wang2016automatically} & Serialized ASTs & Distributed & Defect Prediction\\
        \citet{white2016deep}$^*$ & Token \& Syntax & Distributed & Clone Detection\\ \bottomrule
    \end{tabular}
  \end{center}
\end{minipage}
\end{table}

\section{Applications}
\label{sec:applications}

Probabilistic models of source code have found a wide range of applications
in software engineering and programming language research. 
These models enable the principled use of probabilistic reasoning to
handle uncertainty.  Common sources of uncertainty are
underspecified or inherently ambiguous data (such as natural language text).  In some domains,
probabilistic source code models also simplify or accelerate analysis tasks
that would otherwise be too computationally costly to execute.
In this section, our goal is to explain the use of probabilistic models in each area, not
review them in detail. 
We describe each area's goals and key problems,
then explain how they can benefit from probabilistic, machine learning-based methods,
and how the methods are evaluated.

\subsection{Recommender Systems}
\label{subsec:recomSystems}
Software engineering recommender systems \citep{robillard2010recommendation,robillard2014recommendation}
make recommendations to assist software
engineering tasks, such as code autocompletion and recommending likely code reviewers for
a given code change.
Many of these systems employ data mining and machine learning
approaches on various software engineering artifacts.
Probabilistic models of code find application in source code-based
recommender systems \citep{mens2014source}, such as those that aid developers
write or maintain code.

Modeling developer \emph{intent}
is a challenge: even if there were an agreed upon way to formalize intent, 
developers are reluctant to formalize their intent separately from their
code itself. 
Probabilistic reasoning is well-suited
for inferring intent, since it allows us to quantify the uncertainty that is
inherent to inferring any latent variable.
Probabilistic recommender systems
extract information from the \emph{context} of (partial) code 
and use it to probabilistically reason
about developer intent. 

The most prominent recommender system and a feature commonly used in integrated development environment (IDEs) is
\emph{code completion}. All widely used IDEs, such as Eclipse, IntelliJ and
Visual Studio, have some code completion features. According to \citet{amann2016study},
code completion is the most used IDE feature. However,
code completion tools typically return suggestions in alphabetic order, 
rather than in relative order of predicted relevance to the context. 
Statistical code completion aims to improve suggestion accuracy by learning
probabilities over the suggestions and providing to the users a
ranked list. Some systems focus on automatically completing specific constructs
(\eg method calls and parameters); others try to complete all code tokens.
In all cases, probabilistic code completion systems use existing code as their
training set.

Statistical code completion was first studied by
\citet{bruch2009learning} who extracted features from code context 
to suggest completions for method invocations and constructors. Later, \citet{proksch2015intelligent}
used Bayesian graphical models (structured prediction) to improve 
accuracy. This context-based model captures all usages of an object
and models the probability distribution for the next call. A version of this research is
integrated into the Eclipse IDE under Eclipse Recommenders~\citep{eclipseCodeRecommenders}.

Source code language models have implicitly and explicitly been used
for code completion. \citet{hindle2012naturalness} were the first to
use a token-level \ngram LM for this purpose, using the previous $n-1$ tokens to
represent the completion context at each location. Later, \citet{tu2014localness,franks2015cacheca}
used a cache \ngram LM and further improved the completion performance,
showing that a local cache acts as a domain adapted \ngram.
\citet{nguyen2013statistical} augment the completion context with semantic
information, improving the code completion accuracy of the \ngram LM.
\citet{raychev2014code} exploit formal properties of the code in context to limit
incorrect (but statistically probable) API call suggestions. Their method is
the first to depart from simple statistical token completion towards
statistical program synthesis of single statements.
Apart from token-level language models for code completion, \citet{bielik2016phog} and
\citet{maddison2014structured} create AST-level LMs that can
be used for suggestion.

In contrast to work that predicts source code,
\citet{movshovitz2013natural} create a recommender system to assist comment completion given
a source code snippet, using a topic-like graphical model to model context information.
Similarly, the work of \citet{allamanis2014learning,allamanis2015suggesting,allamanis2016convolutional}
can be seen as a recommender systems for suggesting names for variables, methods, and classes by using relevant code tokens as the context.

\subsection{Inferring Coding Conventions}

Coding conventions are syntactic constraints on code beyond those imposed by
the grammar of a programming language. They govern choices like formatting
(brace and newline placement) or Hungarian notation \vs CamelCase naming.
They seek to prevent some classes of bugs and make 
code easier to comprehend, 
navigate, and maintain~\citep{allamanis2014learning}.
In massive, open, online
courses, coding conventions help teachers identify and understand common student errors \citep{glassman2015overcode}. 
Enforcing coding conventions is tedious.
Worse, it is sometimes difficult to achieve consensus on what they should be, 
a prerequisite for their codification in rule-based systems.
Inferring coding conventions with machine learning solves this problem by \emph{learning}
emergent conventions directly from a codebase.
This can help software teams to determine the coding conventions a codebase uses 
without the need to define rules rule upfront or configure existing convention enforcing tools.

Machine learning models of source code that look at the surface structure (\eg tokens,
syntax) are inherently well-suited for this task. Using the source code as
data, they can infer the emergent conventions while quantifying uncertainty
over those decisions. An important challenge in this application domain is
the sparsity of the code constructs, caused by the diverse and non-repeatable form of
source code within projects and domains.
\citet{allamanis2014learning,allamanis2015suggesting,allamanis2016convolutional,bavishi2017context2name}
exploit the statistical similarities of code's surface structure
to learn and suggest variable, method, and class naming conventions,
while \citet{allamanis2014mining} and \citet{allamanis2016mining} mine conventional syntactic 
and semantic patterns of code constructs that they call \emph{idioms}. They show
that these idioms are useful for documentation and can help software engineering
tool designers achieve better coverage of their tools.
To handle code formatting conventions, 
\citet{parr2016technical} learn a source code
formatter from data by using a set of hand-crafted features from the AST
and a $k$-NN classifier.

\subsection{Code Defects}

Probabilistic models of source code assign high probability to code
that appears often in practice, \ie is natural. Therefore, code considered
very improbable may be buggy. This is analogous 
to anomaly detection using machine learning~\citep{chandola2009anomaly}.
Finding defects is a core problem in software engineering and programming
language research. The challenge in this domain rests in correctly
characterizing source code that contains a defects with high precision and
recall. This is especially difficult because of the rarity of defects and the extreme diversity
of (correct) source code.

Preliminary work suggests that the probability assigned by language models
can indicate code defects. \citet{allamanis2013mining}
suggest that \ngram LMs can be seen as complexity measures and
\citet{ray2015naturalness} present evidence that buggy code tends to have lower probability
(is less ``natural'') than correct code and show that LMs
find defects as well as popular tools like FindBugs.

\citet{wang2016automatically} use deep belief networks to automatically learn
token-level source code features that predict code defects. \citet{fast2014emergent} and
\citet{hsiao2014using}
learn statistics from large numbers of code to detect potentially erroneous
code and perform program analyses while \citet{wang2016bugram} learn coarse-grained
\ngram language models to detect uncommon usages of code. 
These models implicitly assume that a simple set of statistics or an LM can capture 
anomalous/unusual contexts.
Recently, \citet{murali2017finding}
use a combination of topic models to bias a recurrent neural network that models
the sequences of API calls in a learned probabilistic automaton. They use the
model to detect highly improbable sequences of API calls detecting
real-world bugs in Android code. \citet{allamanis2018learning,pradel2017deep}
use various elements from code context to detect specific kinds of bugs, such as variable
and operator misuses.

Because of the sparsity of source code, work on detecting code defects
uses different abstraction levels of source code. For example, 
\citet{wang2016bugram} create coarse-grained \ngrams, while \citet{murali2017finding}
focus on possible paths (that remove control flow dependencies) over API calls.
Therefore, each model captures a limited family of defects, determined by the
model designers' choice of abstraction to represent.
\citet{pu2016skp} and \citet{gupta2017deepfix}
create models for detecting and fixing defects but only for student submissions
where data sparsity is not a problem.
Other data-mining based methods (\eg \citet{wasylkowski2007detecting}) also exist, but are out-of-scope
from this review since they do not employ probabilistic methods.

Also related is the work of \citet{campbell2014syntax} and \citet{bhatia2016automated}.
These researchers use source code LMs to identify and correct syntax errors. Detecting
syntax errors is an easier and more well defined task. The goal of these
models is \emph{not} to detect the existence of such an error (that can be
deterministically found) but to efficiently localize the error and suggest a
fix.

The earlier work of \citet{liblit2005scalable}, \citet{zheng2006statistical} use traces
for statistical bug isolation. \citet{kremenek2007factor} learn factor
graphs (structured prediction) to model resource-specific bugs by modeling resource usage
specifications. These models use an efficient representation to capture bugs, 
but can fail on interprocedural code that
requires more complex graph representations.
Finally, \citet{patra2016learning} use an LM of
source code to generate input for fuzz testing browsers.

Not all anomalous behavior is a bug (it may simply be rare behavior), but anomalous behavior in often executed code almost certainly
is~\citep{engler2001bugs}.  Thus, probabilistic models of source code seem a
natural fit for finding defective code.  They have not, however, seen much
industrial uptake.  One possible cause is their imprecision.  The vast
diversity of code constructs entails sparsity, from which all anomaly
detection methods suffer. Methods based on probabilistic models are no
exception:  they tend to consider rare, but
\emph{correct}, code anomalous.

\subsection{Code Translation, Copying, and Clones}
\label{subsec:migration}

The success of statistical machine translation (SMT)
among (natural) languages has inspired researchers to 
use machine learning to translate code from one source language (\eg Java)
to another (\eg C\#). Although rule-based rewriting systems
can be (and have been) used, it is tedious to create and  maintain these rules, 
in the face of language evolution.
SMT models are well suited for this
task, although they tend to produce invalid code. To reduce these
errors during translation \citet{karaivanov2014phrase} and \citet{nguyen2015divide}
add semantic constraints to the translation process.

Existing research has applied widely used SMT models for text.
Although these models
learn mappings between different language constructs such as APIs, they have only
been used for translating between programming languages of similar paradigms
and structure (C\# and Java are both object-oriented languages with managed memory).
This is an important limitation; machine learning innovations are required 
to translate between languages of different types (\eg Java to Haskell or assembly to C) or
even languages with different memory management (\eg Java to C). 
Existing per-statement SMT from Java to C does
not track memory allocations and therefore fails to emit memory de-allocations
that C's lack of garbage collection requires. Similarly,
translating object-oriented code to functional languages will require learning the
conceptual differences of the two paradigms while preserving semantics, 
such as learning to translate a loop to a map-reduce functional.
Researchers evaluate translation models by scoring exact matches,
measuring the syntactic or semantic correctness of the translated code, or
using BLEU \citep{papineni2002bleu}.
We discuss this and other measure-related issues in \autoref{subsec:futuredirs}.

Developers often copy code during development.  This practice requires fixups 
to rename variables and handle name collisions; it can also create code clones, 
similar code snippets in different locations of a code base \citep{koschke2007survey}.
\citet{allamanis2017smartpaste} automate naming cleanups after copying;  they
use structured prediction
and distributed representations to adapt/port a pasted snippet's variables
into the target context. Their method probabilistically represents
semantic information about variable use to predict
the correct name adaptations without external information (\eg tests).
Clones may indicate refactoring opportunities (that allow reusing the cloned
code).
\citet{white2016deep} use autoencoders 
and recurrent neural networks \citep{goodfellow2016deep} to find
clones as code snippets 
that share similar distributed representations. Using distributed 
vector representations allows them to learn a continuous similarity metric
between code locations, rather than using edit distance.

\subsection{Code to Text and Text to Code}
\label{sec:applications:SourceCodeAndNaturalLanguage}

Linking natural language text to source code has many useful applications, such
as program synthesis, traceability, search and documentation. However, the
diversity of both text and code, the ambiguity of text, the compositional nature of code
and the layered abstractions in software make interconnecting
text and code a hard problem. Probabilistic machine learning
models provide a principled method for 
modeling and resolving ambiguities in text and in code.

Generating natural language from source code,
\ie code-to-text,
has applications to code documentation
and readability.
For example,
\citet{oda2015learning} translate Python code to pseudocode (in natural language) using machine
translation techniques, with the goal of producing
a more readable generation of the code. 
\citet{iyer2016summarizing} design a neural
attention model that summarizes code as text.  
\citet{movshovitz2013natural} generate comments
from code using \ngram models and topic models. 

The reverse direction, text-to-code, aims to help
people, both developers and end users, write
programs more easily.
This area is closely related to semantic parsing in NLP.
Semantic parsing is the task of converting a natural language utterance into a
representation of its meaning, often
database or logical queries that could subsequently be used
for question answering \citep{jurafsky2000speech}.
We do not have space to fully describe the large body of work that has been done in semantic
parsing in NLP, but instead will focus
on text-to-code methods that output code in languages
that are used by human software developers.
This area has attracted growing interest, 
with applications such
as converting natural language to Excel macros \citep{gulwani2014nlyze}, Java expressions \citep{gvero2015synthesizing},
shell commands \citep{lin2017program,lin2018nl2bash}, simple if-then programs \citep{quirk2015language},
regular expressions \citep{kushman2013using} and to SQL queries~\citep{zhong2017seq2sql}.
Finally, \citet{yin2017syntactic} have recently
presented a neural architecture for general-purpose code generation. 
For more details, \citet{neubig2016survey} provides an informal survey of code generation
methods.

\subsection{Documentation, Traceability and Information Retrieval}

Improving documentation and code search are central
questions in software engineering. Probabilistic
models are particularly natural here because, as
we have seen, they allow integrating information
between NL text and code.
Although the more general code-to-text and text-to-code models from the previous sections could
clearly be applied here, researchers
have often found, as in the NL domain, that more
specialized solutions are currently effective
for these problems.

Code search --- a common activity for software engineers \citep{sadowski2015developers,amann2016study} ---
can employ natural language queries.
Software engineering researchers have focused on the code search problem
using information retrieval (IR) methods~\citep{gallardo2009internet,holmes2005strathcona,thummalapenta2007parseweb,mcmillan2013portfolio}.
\citet{niu2016learning} has used learning-to-rank methods but with
manually extracted features. Within the area of statistical models
of source code, \citet{gu2016deep} train a sequence-to-sequence (seq2seq) neural network to map natural
language into API sequences. \citet{allamanis2015bimodal} learn a bimodal, generative
model of code, conditioned on natural language text and use it to rank code search
results. All of these methods use rank-based measures (\eg mean
reciprocal rank) to evaluate their performance.

Documentation is text that captures requirements, specifications, and
descriptions of code.  Engineers turn to it to prioritize new features and to
understand code during maintenance.
Searching, formalizing, reasoning about, and interlinking code to 
(\ie the traceability problem of \citet{gotel2012traceability}) documentation are seminal software engineering 
problems. Mining common
API patterns is a recurring theme and there is a large literature of non-probabilistic methods (\eg
frequency-based) for mining and synthesizing API patterns
\citep{buse2012synthesizing,xie2006mapo}, which are out-of-scope of this review.
Also out-of-scope is work that combines natural language information
with APIs. For example, \citet{treude2016augmenting} extract phrases from
StackOverflow using heuristics (manually selected regular expressions) and
use off-the-self classifiers on a set of hand-crafted features. We refer
the reader to \citet{robillard2010recommendation} for all probabilistic
and non-probabilistic recommender systems.
Within this domain, there are a few probabilistic code models that mine API sequences.
\citet{gu2016deep} map natural language text to commonly used API sequences,
\citet{allamanis2014mining} learn fine-grained source code idioms, that
may include APIs. \citet{fowkes2015parameter} uses a graphical model to mine
interesting API sequences.

Documentation is also related to information extraction from (potentially
unstructured) documents. \citet{cerulo2015irish} use a language model to
detect code ``islands'' in free text. \citet{sharma2015nirmal} use a language
model over tweets to identify software-relevant tweets.

\subsection{Program Synthesis}
Program synthesis is concerned with generating full or partial programs from a specification~\citep{gulwani2017program}.
Traditionally, a specification is a formal statement in an appropriate
logic. More recently, researchers have considered partial or incomplete
specifications, such as input/output pairs \citep{lau2001programming} or a natural language description.
Program synthesis generates full or partial programs from 
a specification. When the specification is a natural
language description,  this is the 
semantic parsing task (see \autoref{sec:applications:SourceCodeAndNaturalLanguage}).
Program synthesis (\eg from examples or a specification) has received a great deal of attention in programming
language research. The core challenge is searching the vast
space of possible programs to find one that complies with the specification.
Probabilistic machine learning models help 
guiding the search process to more probable programs.

Research on programming by example (PBE) leverages machine learning methods
to synthesize code. \citet{liang2010learning} use a graphical model to
learn commonalities of programs across similar tasks with the aim to
improve program synthesis search.
\citet{menon2013machine} use features from the input/output examples to learn a parameterized
PCFG to speed up synthesis. \citet{singh2015predicting}
extract features from the synthesized program to learn a supervised classifier
that can predict the correct program and use it to re-rank synthesis suggestions.
The recent work of \citet{balog2016deepcoder} and 
\citet{parisotto2016neuro} combine ideas from existing enumerative
search techniques with learned heuristics to learn to efficiently 
synthesize code, usually written within a DSL. \citet{neelakantan2015neural}
and \citet{reed2015neural} introduce neural differentiable architectures
for program induction. 
This is an interesting emerging area of research~\citep{kant2018recent}, but does not yet scale to the
types of problems considered by the programming language and software engineering community 
\citep{gaunt2016terpret,feser2017neural}; also see \citet{gaunt2016terpret,robustfill} for a comparison of neural
program induction and program synthesis methods.
Finally, the code completion work of \citet{raychev2014code} can be seen
as a limited program synthesis of method invocations at specific locations.

Although program synthesis is usually referred in the context of generating
a program that complies to some form of specification, probabilistic models
have been used to synthesize random --- but functioning --- programs for benchmarks
and compiler fuzzing. \citet{cummins2017synthesizing}
synthesize automatically a large number of OpenCL benchmarks by learning a character-level LSTM over valid
OpenCL code. Their goal is to generate reasonable-looking
code, rather than synthesize a program that complies with a specification.
To ease their task, they normalize the code by consistently
alpha-renaming variables and method names. Finally, 
they filter invalid intermediate output so they, in the end, 
generate only valid programs. In a similar manner, \citet{patra2016learning}
synthesize JavaScript programs for fuzz testing JavaScript interpreters.

\subsection{Program Analysis}
\label{subsec:progAnalysis}
Program analysis is an important area that seeks to soundly extract semantic
properties, like correctness, from programs.  Sound analyses, especially those
that scale, can be imprecise and return unacceptably large numbers of false
positives.  Probabilistic models of code use probabilistic reasoning to
alleviate these problems.  Three distinct program analysis approaches have
exploited probabilistic models of source code. 

First, a family of models relaxes
the soundness requirement, yielding probabilistic results instead.
\citet{raychev2015predicting} use a graphical model to predict the probability
distribution of JavaScript variable types by learning statistical patterns of how code
is used.
\citet{oh2015learning} and \citet{mangal2015user} use machine learning models
to statistically parameterize program analyses to reduce false positive ratio while maintaining
high precision. \citet{chae2016automatically} reduce automatically (without
machine learning) a program to a set of data-flow graphs, manually extract features
from them. Using these features they then learn the appropriate parametrization
of a static analysis, using a traditional classifier. \citet{koc2017learning}
train a classifier, using LSTMs, to predict if a static analysis warning is
a false positive. The neural network learns common patterns that can discriminate
between false and true positives.
The second paradigm that has been explored is to use machine learning
to create models that produce plausible hypotheses of formal verification statements that 
can be proved. \citet{brockschmidt2017learning} and \citet{li2015gated} propose a set of
models that generate separation logic expressions from the heap graph
of a program, suitable for formally verifying its correctness.
The third paradigm --- although not yet applied directly to source code but
to other forms of formal reasoning --- 
learns heuristics to speed-up the search for a formal automated proof. The
goal of such methods is to replace hard-coded heuristics with a learnable
and adaptive module that can prioritize search tactics per-problem without human
intervention.
\citet{alemi2016deepmath} and \citet{loos2017deep} take a first step towards this
direction by learning heuristic for automated theorem proving for mathematical
expressions.

\section{Challenges and Future Directions}
\label{subsec:futuredirs}

The development and analysis of code must contend with uncertainty,
in many forms:  ``\emph{What is the purpose of this code?}'', ``\emph{What does
functionality does this unit test test?}'',  ``\emph{From this program, can we infer
any of its specification?}'', ``\emph{What is this program's intended (or likely) input
domain?}'' or ``\emph{Why did this program crash here?}''.
This 
contrasts with traditional program analysis which is conservative: it deems that a program
has a property, like a bug, if that property is possible, independent of likelihood.
In contrast, machine learning studies robust inference under uncertainty and noise.
Thus, the application of machine learning to code and its development, 
is an emerging research topic with the potential to influence
programming language and software engineering research.
Here, we list topics where principled probabilistic reasoning promises new
advances, focusing on probabilistic models of code.
We also list open challenges, some quite longstanding like long-range
dependency, and speculate about directions for making progress on their
resolution.

For each open problem, machine learning is introduced to handle uncertainty,
ambiguity, or avoid hard-coded heuristics. 
Probabilistic learning systems model these as noise which they handle 
robustly by training statistical principled models.
This in turn has allowed the creation of new previously impossible systems
(\eg text-to-code systems)
or replaced existing, hard-coded heuristics with
machine learning systems that promise to be
more robust and generalize better.
This course of direction highly resembles
that of NLP, where hard-coded ``expert'' systems have been successfully
replaced by sophisticated machine learning methods.

\subsection{The Third Wave of Machine Learning}
\label{subsec:mlFd}

The first wave
of machine learning for source code applied off-the-shelf
machine learning tools with hand-extracted features. The second wave,
reviewed here, avoids manual feature extraction and uses the source code
itself within machine learning heavily drawing inspiration from existing
machine learning methods in NLP and elsewhere. The third wave promises
new machine learning models informed by programming language semantics.
What form will it take?

At the time of this writing, machine learning, and deep learning in particular,
is enjoying rock-star status among research fields. Despite its current
(perhaps ephemeral) popularity, it is not a panacea.  In some cases, a machine
learning model may not be required (\eg when the problem is deterministic) and,
in other cases, simple models can outperform advanced, off-the-self deep
learning methods, designed for non-code domains~\citep{fu2017easy,hellendoorn2017deep}.
Furthermore, over-engineering or under-engineering machine learning models
usually leads to suboptimal results. Selecting a machine learning model for a
specific problem necessitates questioning if a specific model is fit for the
target application.  Strong baseline methods are needed to estimate if a
performance improvement justifies the added complexity of a model.  In short,
the right tools should be used for the right job, and machine learning is no
exception. 

\paragraph{Bridging Representations and Communities}
 Programming language research and practice use a
well-defined and widely useful set of representations, usually in a
symbolic form. In contrast, machine learning research customarily works
with continuous representations. Bridging the gap between these representations
by allowing machine learning systems to reason with programming language
representations, is an important challenge. Handling unambiguous code has
already led to a combination of probabilistic methods with ``formal''
constraints, that limit the probabilistic model to valid code. For example,
\citet{maddison2014structured} limit their model to generate syntactically
correct and scope-respecting code while \citet{raychev2015predicting}
easily create a highly-structured model of a program taking advantage of its 
unambiguous form. Introducing better representations that bridge the gap
between machine learning and source code will allow the probabilistic models of code to reason about the rich
structure and semantics of code, without resorting to proxies. The core
problem in this area is the lack of understanding of machine learning from
the programming language community and vice-versa. At the same time, new
machine learning methods that can handle programming language structures in
its full complexity at scale need to be researched.

A major obstacle here is engineering systems that efficiently and effectively
combine the probabilistic world of machine learning and the formal, logic-based
world of code analysis. One approach, taken by
several authors \citep{allamanis2017learning,raychev2015predicting,rocktaschel2017end,mangal2015user}, is to relax formal systems into probabilistic. 
Such systems, however, lack the guarantees formal systems (\eg soundness) often
provide.
\citet{alemi2016deepmath}, \citet{balog2016deepcoder}
and \citet{loos2017deep} follow a second approach that maintains soundness:  they
learn input and context-specific heuristics that efficiently guide search-based methods,
such as theorem proving and program synthesis. 

All of approaches to source code modeling must decide whether to explicitly
model the structure and constraints of source code or to rely on 
general methods with adequate capacity. One one hand, using well-known, domain-generic
machine learning methods has the advantage that the models are well-understood 
and rarely require significant effort or expertise to apply.
On the other hand, designing models with built-in 
inductive biases for the problem domain usually performs better with less
data, at the cost of manually designing and debugging domain (even problem-specific) networks.
One such promising direction is modular neural network architectures.
Such architectures decompose the network
into components that are combined based on the problem instance.
For source code models, such architectures can derive its structure through static analyses.
These architectures have been useful for visual question answering in NLP. 
\citet{andreas2016learning} create neural networks by composing them from ``neural modules'',
based on the input query structure.
Similarly, we believe that such architectures will be useful within probabilistic
models of source code.
An early example is the work of \citet{allamanis2018learning} who design a
neural network based on the output of data flow analysis. Such architectures
should not only be effective for bridging representations among communities but
--- as we will discuss next --- can combat issues with compositionality,
sparsity and generalization. Nevertheless, issues that arise in static analyses,
such as path explosion, will still need to be addressed.

\paragraph{Data Sparsity, Compositionality and Strong Generalization} 
The principle of reusability in software
engineering creates a form of sparsity in the data, where it is rare to find multiple
source code elements that perform exactly the same tasks. For example, it is
rare to find hundreds of database systems, whereas one can easily find thousands
of news articles on a popular piece of news.  
The exceptions, like programming competitions and student solutions to programming assignments,
are quite different from industrial code.  This suggests that there are many
opportunities for researching machine learning models and inference methods that can handle and
generalize from the highly-structured, sparse and composable nature of source code data.
Do we believe in the unreasonable effectiveness of data~\citep{halevy2009unreasonable}?
Yes, but we do not have sufficient data.

Although code and text are both intrinsically extensible,
code pushes the limit of existing machine learning methods in terms
of representing composition. This is because
most natural language methods rarely define novel, previously unseen, terms, 
with the possible exception of legal and scientific texts. In
contrast, source code is inherently extensible, with developers constantly
creating new terms (\eg by defining new functions and classes) and combining
them in still higher-level concepts. Compositionality refers to the idea that
the meaning of some element can be understood by composing the meaning of its
constituent parts. Recent work \citep{hill2016learning}
has shown that deep learning architecture can learn some aspects of compositionality in text.
Machine learning for highly compositional objects remains challenging, because it has
proven hard to capture relations between objects, especially across abstraction levels.
Such challenges arise even when considering simple code-like expressions
\citep{allamanis2017learning}. However, if sufficient progress is
to be made, representing source code artifacts in machine learning will
improve significantly, positively affecting other downstream tasks.
For example, learning composable models that can combine meaningful representations of
variables into meaningful representations of expressions and functions
will lead to much stronger generalization performance.

Data sparsity is still an important and unsolved problem. Although finding a reasonably large amount of source code is
relatively easy, it is increasingly hard to retrieve some representations
of source code. Indeed, it is infeasible even to compile all of the projects
in a corpus of thousands of projects, because compiling a project requires
understanding how the project handles external dependencies, which can sometimes be idiosyncratic.
Furthermore, computing or acquiring semantic properties of 
existing, real-world code (\eg 
purity of a function~\citep{finifter2008verifiable} or pre-/post- conditions~\citep{Hoare:1969:ABC:363235.363259}) is
hard to do, especially at scale.  Scalability also hampers harvesting 
run-time data from real-world programs: it is challenging to acquire substantial run-time data even for 
a single project. Exploring ways to synthesize or transform ``natural'' programs that perform the same task in different
ways is a possible way ahead. Another promising direction to
tackle this issue is by learning to extrapolate from run-time data (\eg
collected via instrumentation of a test-suite) to static properties of
the code \citep{allamanis2016mining}. Although this is inherently a noisy process, achieving high
accuracy may be sufficient, thanks to the inherent ability of machine learning
to handle small amounts of noise.

Strong generalization also manifests as a deployability problem.
Machine learning models, especially when they have become effective, are often so large that they are
too large for a developer's machine, but using the cloud raises 
privacy\footnote{\url{https://www.theregister.co.uk/2017/07/25/kite_flies_into_a_fork/}} concerns and prevents
offline coding. When under development and 
tooling is needed, code evolves quickly, subjecting models to constant concept drift and
necessitating frequent retraining which can be extremely slow and costly.  Addressing this
deployability concern is an open problem and requires advances in machine learning areas
such as transfer learning and one-shot learning. For example, say a program $P$ uses libraries $A$ and $B$, which have
been shipped with the models $M_A$ and $M_B$.  Could we save time training a model for
$P$ by transferring knowledge from $M_A$ and $M_B$?

Finally, source code representations are multifaceted. For example, the token-level
``view'' of  source code is quite different from a data flow view of code. Learning
to exploit multiple views simultaneously can help machine learning models generalize
and tackle issues with data sparsity. Multi-view \citep{xu2013survey}
and multi-modal learning (\eg \citet{gella2016unsupervised}), areas actively explored in machine learning, aim
to achieve exactly this. By combining multiple representations of data, they aim
to improve upon the performance on various tasks, learning to generalize using
multiple input signals. We believe that this is a promising future direction
that may allow us to combine probabilistic representations of code to achieve
better generalization.

\paragraph{Measures}

To train and evaluate machine learning models, we need to easily measure their
performance. These measures allow the direct comparison of models and have already lead to
improvements in multiple areas, such as code completion (\autoref{subsec:recomSystems}).
Nonetheless, these measures are imprecise.  For instance, probabilistic recommender
systems define a probability density over suggestions whose cross-entropy can
be computed against the empirical distribution in test data.  Although cross-entropy
is correlated with suggestion accuracy and confidence, small
improvements in cross entropy may not improve accuracy.  Sometimes the
imprecision is due to unrealistic use case assumptions.  For example, the
measures for LM-based code completion tend to assume that code is written
sequentially, from the first token to the last one. However, developers rarely
write code in such a simple and consistent way \citep{proksch2016evaluating}.
Context-based approaches assume that the available context (\eg other object
usages in the context) is abundant, which is not true in a real editing
scenarios.  Researchers reporting keystrokes saved have usually assumed that
code completion suggestions are continuously presented to the user as she is
typing. When the top suggestion is the target token, the user presses a single
key (\eg return) to complete the rest of the target.

Furthermore, some metrics that are widely used in NLP are not suited for source code.
For example, BLEU score is not suitable for measuring the quality of output source code (\eg
in transducer models) because it fails to account for the fact that the 
syntax is known in all programming languages, so the BLEU score may be
artificially ``inflated'' for predicting deterministic syntax. Second, the granularity over which
BLEU is computed (\eg per-statement \vs per-token) is controversial.
Finally, syntactically diverse answers may be semantically equivalent,
yielding a low BLEU score while being correct. Finding new widely-accepted
measures for various tasks will allow the community to build reliable models
with well-understood performance characteristics.

\subsection{New Domains}

Here, we visit a number of domains to which machine learning has not yet been
systematically applied and yet suffer from uncertainty problems that machine
learning is particularly well suited to address, promising new advances.

\paragraph{Debugging} Debugging is a common task for software
engineers \citep{wong2016survey}. Debugging is like trying to find a needle in
the haystack; a developer has to recognize relevant information from the deluge
of available information.  A multitude of tools exist in this area whose main
goal is to visualize a program's state during its execution. Probabilistic
models of source code could help developers, such as by filtering
highly-improbable program states. Statistical debugging models,
such as the work of \citet{zheng2003statistical,zheng2006statistical} and
\citet{liblit2005scalable} are indicative of the possibilities within
this area.
Further adding learning within
debugging models may allow further advances in statistical debugging.
However, progress in this area is impeded by
the combination of lack of data at a large scale and the inherent difficulty of
pattern recognition in very high-dimensional spaces. Defects4J \citep{just2014defects4j}
--- a curated corpus of bugs --- could further prove useful within machine learning for
fault prediction. Furthermore, collecting and filtering
execution traces to aid debugging is another challenge for which machine
learning is well-suited.  Collection requires expensive instrumentation, which
can introduce Heisenbugs, bugs masked by the overhead 
of the instrumentation added to localize them.  Here the question is ``Can machine learning identify probe points or
reconstruct more complete traces from  partial traces?''  Concerning filtering
traces, machine learning may be able to find interesting locations, like the 
root cause of bugs. Future methods should also be able to 
generalize across different programs, or even different revisions of the
same program, a difficult task for existing machine learning methods.

\paragraph{Traceability}
Traceability is the study of links among software engineering artifacts.   
Examples include
links that connect code to its specification, the specification to requirements,
and fixes to bug reports.
Developers can exploit these links to better understand and maintain their 
code.  Usually, these links must be recovered.
Information retrieval has dominated link recovery. The work
of \citet{guo2017semantically} and \citet{le2015rclinker}
suggests that learning better (semantic) representations of artifacts can successfully,
automatically solve important traceability problems.

Two major obstacles impede progress: lack of data and a focus on generic 
text.  Tracing discussions in email threads, online chat rooms (\eg Slack), 
documents and source code would be extremely useful, but no
publicly available and annotated data exists. Additionally, to date, NLP research
has mostly focused on modeling generic text (\eg from newspapers); 
technical text in conversational environments (\eg chatbots) has only begun to be researched.
\href{https://stackoverflow.com}{StackOverflow} presents one such interesting
target. Although there are hundreds of studies that extract useful
artifacts (\eg documentation) from StackOverflow, NLP methods --- such as
dependency parsing, co-reference analysis and other linguistic phenomena
--- have not been explored.

\paragraph{Code Completion and Synthesis}
Code completion and synthesis using machine learning are two heavily
researched and interrelated areas. Despite this fact,
to our knowledge, there has been no full scale comparison between LM-based
\citep{hindle2016naturalness,nguyen2013statistical,raychev2014code}
and structured prediction-based autocompletion models \citep{bruch2009learning,proksch2015intelligent}.
Although both types of systems target the same task, the lack of a well-accepted
benchmark, evaluation methodology and metrics has lead to the absence of a
quantitative comparison that highlights the strengths and weaknesses of each
approach. This highlights the necessity of widely accepted, high-quality
benchmarks, shared tasks, and evaluation metrics that can lead to comparable 
and measurable improvements to tasks of interest. 
NLP and computer vision follow such a paradigm with great success\footnote{See
\url{https://qz.com/1034972/} for a popular account of the effect of large-scale datasets
in computer vision.}.

\citet{omar2017toward} discuss the challenges that arise from the fact that
program editors usually deal with incomplete, partial programs.
Although they discuss how formal semantics can extend to these
cases, inherently any reasoning about partial code requires reasoning about 
the programmer's intent. \citet{lu2017data} used information-retrieval methods
for synthesizing code completions showing that simply retrieving snippets from ``big code'' can be
useful when reasoning about code completion, even without a learnable probabilistic
component.  This suggests a fruitful area for probabilistic models
of code that can assist editing tools when reasoning about incomplete code's semantics,
by modeling how code could be completed.

\paragraph{Education} Software engineering education is one area that is
already starting to be affected by this field. The work of \citet{campbell2014syntax}
and \citet{bhatia2016automated} already provide an automated method for fixing
syntax errors in student code, whereas  \citet{wang2017deep,piech2015learning}
suggest advancements towards giving richer feedback to students. Achieving
reasonable automation can help provide high-quality computer science education
to many more students than is feasible today.
However, there are important challenges associated with this area. This includes
the availability of highly-granular data where machine learning systems 
can be trained, difficulty with embedding semantic features of code into
machine learning methods and the hardness of creating models that can generalize
to multiple and new tasks. 
Student coursework submissions are potentially a ripe application
area for machine learning, because here we have available many programs, from
different students, which are meant to perform the same tasks. An especially
large amount of such data is available in Massive Open Online Courses (MOOCs).
This opens exciting possibilities, such as
providing granular and detailed feedback, curriculum customization and other
intelligent tutoring systems can significantly change computer science
education.

\paragraph{Assistive Tools}
Probabilistic models have allowed computer systems to handle noisy inputs
such as speech, and handwritten text input. In the future,
probabilistic models of source code may enable novel
assistive IDEs, creating inclusive tools that improve upon
conventional methods of developer-computer interaction and provide inclusive
coding experiences.

\section{Related Research Areas}
\label{sec:related}

A variety of related research areas within software engineering
and programming languages overlap with the area of statistical modeling of code.
One of the most closely related research areas is \emph{mining software repositories} (MSR) and \emph{big code}.
MSR is a  well-esta\-blished, vibrant and active field; the idea is 
to mine  the large amounts of source code data and meta-data available
in open-source (and commercial) repositories to gain valuable
information, and use this information to enhance both tools  and processes. 
The eponymous flagship conference is now in 
its 15$^{th}$ iteration. ``Big Code'' is a synonymous neologism,
created by DARPA's MUSE program\footnote{\url{http://science.dodlive.mil/2014/03/21/darpas-muse-mining-big-code/}} 
to borrow some branding shine from the well-marketed term \emph{``Big Data''}. 
The MSR field's early successes date back to work by \citet{zimmermann2005mining},
\citet{williams2005automatic}, \citet{gabel2008javert} and \citet{acharya2007mining} on mining API
protocols from source code bases. These approaches used pragmatic counting 
techniques, such as frequent item-set mining.
Mining software repositories is in one sense a broader field than statistical models of code,
as rather than focusing on code alone, MSR considers the full spectrum of software engineering data that can be derived
from the software engineering process, such as process measures,
requirement traceability, commit logs.
Additionally, research in 
malware detection is related to probabilistic models of code~\citep{arp2014drebin,cen2015probabilistic}.

Another active area at the intersection machine learning and programming language research is \emph{probabilistic programming}
\citep{gordon2014probabilistic}. This might appear to be related to statistical models of code, but in fact there is a fundamental difference; essentially, probabilistic programming works in the reverse direction. Probabilistic programming
seeks to deploy programming language concepts to make it easier for developers to write new machine learning algorithms.
Statistical code models seek to apply machine learning concepts to make it easier for developers to write new programs.
In some sense, the two areas are dual to each other. That being said, one can certainly imagine completing the cycle
and attempting to develop statistical code models for probabilistic programming languages. This could be a fascinating
endeavor as probabilistic programming grows in popularity to the extent that large corpora of probabilistic programs
become available.

In software engineering, the term \emph{modeling} often refers to 
formal specifications of program behavior, which are clearly a very different kinds of models than
those described here. Combining formal models of semantics with statistical models of source code 
described in this review would be an interesting area for future research.
There is also some work on probabilistic models of
code that do \emph{not} have a learning component, such as \citet{liblit2005scalable}. 
Within machine learning, there has been an interesting recent line of work on neural 
abstract machines \citep{graves2014neural,reed2015neural,riedel2016programming},
which extend deterministic automata from computer science, such as pushdown automata and Turing machines,
to represent differentiable functions, so that the functions can be estimated by techniques from 
machine learning. To the best of our knowledge, this intriguing line of work does not yet
consider source code, unlike the work described in this review.
Finally, semantic parsing is a vibrant research area of NLP
that is closely related to the idea of program synthesis from natural language; see \autoref{sec:applications:SourceCodeAndNaturalLanguage} for more discussion.

\section{Conclusions}
\label{sec:conclusions}
Probabilistic models of source code have exciting
potential to support new tools in almost every
area of program analysis and software engineering.
We reviewed existing work in the area, presenting a taxonomy of probabilistic machine learning source code models and their applications.
The reader may appreciate that most of the research contained in this review
was conducted within the past few years, indicating a growth of
interest in this area among the machine learning, programming languages and software
engineering communities. Probabilistic models of source code raise the
exciting opportunity of \emph{learning} from existing code,
probabilistically reasoning about new source code artifacts
and transferring knowledge between developers and projects.

\begin{screenonly}	

\end{screenonly}

\appendix

\section{Supplementary materials}

\begin{printonly}

\begin{table}[tb]
   \caption{Some datasets, available online, used in research of Probabilistic Models of Source Code (sorted alphabetically).
      Links are clickable in the digital version.}
      \label{tbl:datasets}
       \begin{minipage}{\columnwidth}\begin{center}\footnotesize
  \begin{tabular}{lp{9cm}l}
        \toprule
        Reference & Short Description & Link \\ \midrule
        \citet{allamanis2013mining} & Deduplicated snapshot of all Java GitHub projects with least one fork. & \href{http://dx.doi.org/10.7488/ds/1690}{link}\\
        \citet{allamanis2016convolutional} & Parsed source code for 11 highly-ranked Java GitHub projects. & \href{http://groups.inf.ed.ac.uk/cup/codeattention/}{link}\\
        \citet{barone2017parallel} & Parallel corpus of 150k Python function declarations, docstrings and bodies. & \href{https://github.com/EdinburghNLP/code-docstring-corpus}{link}\\
        \citet{cerulo2015irish} & Free text data with source code ``islands''. & \href{www.rcost.unisannio.it/cerulo/dataset-scam2013.tgz}{link}\\
        \citet{boa} & 800k+ code repositories, software to support queries and mining & \href{http://boa.cs.iastate.edu/}{link} \\
        \citet{iyer2016summarizing} & Source code snippets with their StackOverflow title.& \href{https://github.com/sriniiyer/codenn/tree/master/data/stackoverflow}{link}\\
        \citet{kushman2013using} & Dataset for generating regular expressions from natural language. & \href{https://groups.csail.mit.edu/rbg/code/regexp/}{link}\\
        \citet{lin2018nl2bash} & Text to Bash commands corpus. & \href{https://github.com/TellinaTool/nl2bash}{link} \\ 
        \citet{ling2016latent} & Text descriptions and code for card game. & \href{https://github.com/deepmind/card2code}{link}\\
        \citet{raychev2016learning} & A dataset of deduplicated JavaScript files and their ASTs extracted from GitHub.& \href{http://www.srl.inf.ethz.ch/js150.php}{link} \\
        \citet{raychev2016learning} & A dataset of deduplicated Python files and their ASTs extracted from GitHub.& \href{http://www.srl.inf.ethz.ch/py150}{link} \\
        \citet{oda2015learning} & Parallel corpus of Python code and pseudocode in English and Japanese. & \href{http://ahclab.naist.jp/pseudogen/}{link} \\\bottomrule
    \end{tabular}
  \end{center}
\end{minipage}
\vspace{-1em}
\end{table}

\end{printonly}

\begin{screenonly}
\autoref{tbl:datasets} presents datasets that have been specifically created for
this area of research.
  
\end{screenonly}


\bibliographystyle{ACM-Reference-Format-Journals}
\bibliography{lit/bibliography}
\end{document}